%% file: usenix.tex
\documentclass[letterpaper,twocolumn,10pt]{article}
\usepackage{usenix,epsfig,endnotes}
\usepackage{color}
\usepackage{graphicx}
\usepackage[available]{usenixbadges}
\usepackage{algorithmic}
\usepackage{algorithm}
\usepackage{xspace}
\usepackage{amsmath}
\usepackage{caption}
\usepackage{subfig}
\usepackage{multirow}
\newcommand{\hawkeye}{\texttt{\texttt{HawkEye}}\xspace}
\newcommand{\mpspdz}{\texttt{\texttt{MP-SPDZ}}\xspace}

\newcommand{\share}[1]{\langle {#1} \rangle}
\usepackage{amsmath}
\usepackage{amssymb}
\usepackage{listings}
\newcommand{\wqruan}[1]{\textcolor{black}{#1}}
\newcommand{\wqruanother}[1]{\textcolor{black}{#1}}

\DeclareCaptionFormat*{mystyle}{\noindent\hrulefill\par\bfseries#1.\space\normalfont#3\par}
\begin{document}

%don't want date printed
\date{}

%make title bold and 14 pt font (Latex default is non-bold, 16 pt)
\title{\hawkeye: Statically and Accurately Profiling \\the  Communication Cost of Models in Multi-party Learning}

\author{
{\rm Wenqiang Ruan}\\
Fudan University
\and
{\rm Xin Lin}\\
Fudan University
\and
{\rm Ruisheng Zhou}\\
Fudan University
\and
{\rm Guopeng Lin}\\
Fudan University
\and
{\rm Shui Yu}\\
University of Technology Sydney
\and
{\rm Weili Han\thanks{Corresponding author: wlhan@fudan.edu.cn}}\\
Fudan University
}

\maketitle

% Use the following at camera-ready time to suppress page numbers.
% Comment it out when you first submit the paper for review.
% \thispagestyle{empty}

\begin{abstract}
Multi-party computation (MPC) based machine learning, referred to as multi-party learning (MPL), has become an important technology for utilizing data from multiple parties with privacy preservation. In recent years, in order to apply MPL in more practical scenarios, various MPC-friendly models have been proposedto reduce the extraordinary communication overhead of MPL. Within the optimization of MPC-friendly models, a critical element to tackle the challenge is profiling the communication cost of models. However, the current solutions mainly depend on manually establishing the profiles to identify communication bottlenecks of models, often involving burdensome human efforts in a monotonous procedure.

In this paper, we propose \texttt{HawkEye}, a static model communication cost profiling framework, which enables model designers to get the accurate communication cost of models in MPL frameworks without dynamically running the secure model training or inference processes on a specific MPL framework. Firstly, to profile the communication cost of models with complex structures, we propose a static communication cost profiling method based on a prefix structure that records the function calling chain during the static analysis. Secondly, \texttt{HawkEye} employs an automatic differentiation library to assist model designers in profiling the communication cost of models in \texttt{PyTorch}. Finally, we compare the static profiling results of \texttt{HawkEye} against the profiling results obtained through dynamically running secure model training and inference processes on five popular MPL frameworks, \texttt{CryptFlow2}, \texttt{CrypTen}, \texttt{Delphi}, \texttt{Cheetah}, and \texttt{SecretFlow-SEMI2K}. The experimental results show that \texttt{HawkEye} can accurately profile the model communication cost without dynamic profiling. 
\end{abstract}
\input{tex/1.introduction}

\input{tex/2.preliminaries}

\input{tex/3.Cost_analysis}

\input{tex/4.Autograd}

\input{tex/5.Experiments}
\input{tex/6.RelatedWork}

\input{tex/7.Discussion}

\input{tex/8.Conclusion}

\section{Ethics considerations} 
We analyze the ethics of our paper from the following three principles: beneficence, respect for persons, and justice. Regarding beneficence, because we do not test any live services or APIs that give access to otherwise non-public algorithms or models, our results would not cause any financial loss. Regarding respect for persons, our study aims to improve the efficiency of MPC-friendly model development and does not have risks of disrespecting persons. Regarding justice, the results of our study would expand the practical applications of MPL and not impact any special stakeholder group.

\section{Compliance with the open science policy}

We have attended the artifact evaluation and open-source our codes in \url{https://zenodo.org/records/14764461}.

\section*{Acknowledgment}
This paper is supported by Natural Science Foundation of China (62172100, 92370120) and the National Key R\&D Program of China (2023YFC3304400). We thank Lei Kang, Xinyu Tu, Cheng Hong, Wen-jie Lu, and all anonymous reviewers for their insightful comments.

\bibliographystyle{acm}

\appendix

\input{tex/9.appendix}

% \theendnotes

\end{document}

%% file: tex/1.introduction.tex
\section{Introduction}~\label{sec:intro}
As more and more countries publish privacy protection regulations, such as GDPR from the EU, multi-party computation (MPC) based machine learning, referred to as multi-party learning (MPL)~\cite{song2020sok,10.1145/3548606.3560697}, has become an important technology for utilizing data from multiple parties with privacy preservation. On the one hand, in \wqruanother{industry}, many IT giants have released MPL frameworks, such as \texttt{CrypTen}~\cite{crypten2020} released by Facebook, \texttt{TF-Encrypted}~\cite{TF-Encrypted} released by Google, and \texttt{SecretFlow}~\cite{secretflow} released by Ant group et al., to meet their data sharing requirements with high-security restrictions, where raw data must be stored distributively. For example, Ant group applies \texttt{SecretFlow} to deliver personalized policies for different customers of insurers. On the other hand, in \wqruanother{academia}, researchers have proposed dozens of MPL frameworks, such as \texttt{Cheetah}~\cite{Cheetah}, \texttt{Delphi}~\cite{244032},  and \texttt{CRYPTGPU}~\cite{cryptGPU}, in recent years. However, although MPL has \wqruanother{started to be used broadly}, the huge communication overhead still seriously limits its practical applications.

In recent years, \wqruanother{to reduce the huge communication overhead of MPL, many MPC-friendly models~\cite{ganesan2022efficient, MPCViT, li2023mpcformer, liu2023llms, Peng_2023_ICCV, dhyani2023privit} have been proposed.} \wqruanother{Within} the optimization of MPC-friendly models, a critical element to \wqruanother{tackle} the challenge is profiling the communication cost of models. For example, through profiling the communication cost of the secure DenseNet-121~\cite{Huang_2017_CVPR} inference process, Ganesan et al.~\cite{ganesan2022efficient} \wqruanother{found} that the convolution layer is the communication bottleneck. Then, they \wqruanother{designed} an optimized convolution operator to reduce its communication overhead. \wqruanother{In} the above process, profiling the model communication cost is a key step, which is also a popular methodology of efficiency optimization in the machine learning field~\cite{10.1145/3379337.3415890, NEURIPS2022_4fc81f4c}.

Even though model communication cost profiling is a key step in the design of MPC-friendly models, an effective model communication cost framework is still absent. Therefore, model designers have to manually establish the profiles to find communication bottlenecks of models, often involving burdensome human efforts in a monotonous procedure.
Intuitively, model designers can profile model communication cost by inserting test instruments between each layer and dynamically running the secure model training or inference processes on a specific MPL framework~\cite{ganesan2022efficient, MPCViT, li2023mpcformer, dhyani2023privit}. We refer to this method as dynamic profiling. Although dynamic profiling can accurately find the communication bottleneck of models, it requires specific designs and implementation for different models. As a result, dynamic profiling usually takes burdensome human efforts and computing resources, thus being less efficient. On the other hand, a few MPL frameworks, such as \texttt{SecretFlow-SPU}~\cite{secretflow}, support model communication cost profiling at the operation level, \wqruanother{i.e.,} outputting the communication cost of each basic operation (\wqruanother{e.g.,} multiplication, comparison). \wqruanother{However, because \texttt{SecretFlow-SPU} uses Google's XLA~\cite{50530} as a black box to compile the input program, it requires model designers to manually construct many models with a single layer (\wqruanother{e.g.,} convolution) to test the communication cost of different model components.} Note that it is the key to analyzing the communication cost of each layer when model designers try optimizing MPC-friendly model~\cite{ghodsi2020cryptonas}.

\wqruanother{In this paper, we propose \hawkeye, a static model communication cost profiling framework to accurately profile model communication cost without dynamic profiling.} Firstly, to profile the communication cost of models with complex structures, we propose a static communication cost profiling method based on a prefix structure that records the function calling chain. We describe the details of the prefix structure in Section~\ref{subsec:block_label_analysis}. With this method, the communication cost of models can be automatically profiled without manually inserting test instruments and dynamically running the secure model training or inference process. Because the program execution flow must be input-independent to ensure security in MPL frameworks, we can accurately know the computations required by a secure model training or inference process through such static analysis. Therefore, the static profiling results from \hawkeye are consistent with those obtained by the dynamic profiling. 

Secondly, \hawkeye employs an automatic differentiation (\wqruanother{Autograd}) library, which integrates our proposed static communication cost profiling method, to assist model designers in profiling the communication cost of models in \texttt{PyTorch}. In particular, to avoid the bias brought by the extra communication overhead of the broadcast operator, which is commonly used to construct machine learning models, we design an optimized broadcast operator that is communication-efficient for various MPL frameworks. Based on our proposed static communication cost profiling method and the Autograd library, \hawkeye can receive \texttt{PyTorch}-based model training or inference codes, then automatically profile model communication cost. As a result, with \hawkeye, model designers can focus on optimizing the structure of MPC-friendly models without spending much time on profiling model communication costs.

Finally, we compare the static profiling results outputted by \hawkeye with the dynamic profiling results obtained from \wqruanother{five MPL frameworks,  \texttt{CryptFlow2}~\cite{rathee2020cryptflow2}, \texttt{CrypTen}~\cite{crypten2020}, \texttt{Delphi}~\cite{mishra2020delphi}, \texttt{Cheetah}~\cite{Cheetah}, and \texttt{SecretFlow-SEMI2K}~\cite{secretflow}}. The experimental results show that \hawkeye can accurately profile the communication cost of models in various MPL frameworks without dynamic profiling. Furthermore, we conduct \wqruanother{three} case studies to show the practical applications of \hawkeye (Section~\ref{subsec:case_studies}). For example, by applying \hawkeye to profile the communication cost of models in four MPL frameworks with different security models, we find that under the same assumption on the number of colluded parties, MPC-friendly models optimized for MPL frameworks with semi-honest security models remain effective when model designers transfer the same optimization for MPL frameworks with malicious security models. The above results show that  \hawkeye can effectively help model designers optimize the structures of MPC-friendly models without dynamic profiling. 

To summarize, we highlight our contributions as follows.
\begin{itemize}
    \item  We propose a static model communication cost profiling method to accurately analyze the communication cost of models in MPL frameworks without dynamic profiling.
    
    \item  We design and implement \hawkeye with an Autograd library to assist model designers in profiling the communication cost of models in \texttt{PyTorch}. 
    % \wqruanother{Therefore, with \hawkeye, model designers can quickly find the communication cost bottleneck of models run on different MPL frameworks and optimize MPC-friendly models agilely.} 
\end{itemize}

% \wqruanother{We organize the rest of the paper as follows. In Section~\ref{sec:pre} includes the necessary background knowledge to understand the paper.}

%% file: tex/2.Preliminaries.tex
\section{Preliminaries}~\label{sec:pre}
\vspace{-6mm}
\subsection{Multi-Party Learning}~\label{subsec:mpl}
MPL enables multiple parties to collaboratively perform model training or inference on their private data with privacy preservation. Since Mohassel and Zhang published \texttt{SecureML}~\cite{DBLP:conf/sp/MohasselZ17}, the pioneer study of MPL, it has become a significant topic in both \wqruanother{industry and academia}. Currently, There are mainly two technical routes to implement MPL: secret sharing~\cite{ref_damgard} and homomorphic encryption (HE)~\cite{yi2014homomorphic}. Because secret sharing-based MPC protocols usually have higher efficiency in arithmetic operations, MPL frameworks mainly take secret sharing-based MPC protocols as their underlying protocols to implement secure model training or inference.  \wqruanother{HE is typically used to implement secure model inference because it can significantly reduce the computation and communication overhead of clients who hold data.}
% In \hawkeye, we focus on model communication cost profiling on secret sharing-based MPL frameworks~\cite{aby, mohassel2018aby3}, whose performance bottleneck is usually the communication cost.

\smallskip
\noindent\textbf{Fixed-point number representation and computation.} Fixed-point number is a widely used data representation in MPL. A fixed-point number $\widetilde{x}$ with $f$-bit precision is encoded by mapping it to an integer $\overline{x}$, \wqruanother{i.e.,} $\overline{x} = \widetilde{x} * 2^f$, where $\overline{x}$ is an element of ring or field. Since Multiplication would double the fractional part of the output, \wqruanother{i.e.,} $\overline{z} = \overline{x}*\overline{y} = \widetilde{x} * \widetilde{y} * 2^{2f}$, we need to perform truncation on the output to restore the bit length of its fractional part as $f$. 

We then introduce the basic operations that are necessary for a secret sharing-based MPL framework as follows:

   \noindent \textbf{Share}: Given an input $x$, it generates $m$ shares $\share{x}_1, \cdots$, $\share{x}_m$ and distributes them to corresponding parties.
   
   \noindent \textbf{Reveal}: Given $m$ shares $\share{x}_1, \cdots, \share{x}_m$, it reconstructs the original value $x$.
   
    % \item Addition: Given two shares $\share{x}_i, \share{y}_i$, Addition outputs $\share{z}_i$ to $P_i$ such that $z = x+y$.
   \noindent \textbf{Multiplication}: Given two shares $\share{x}_i, \share{y}_i$, it outputs $\share{z}_i$ to $P_i$ such that $z = x*y$. 
    % \item Matrix\_Multiplication: Given two secret shared matrix $\share{\mathbb{X}}_i$, $\share{\mathbb{Y}}_i$, it outputs $\share{\mathbb{Z}}_i$ to $P_i$ such that $\mathbb{Z} = \mathbb{X}*\mathbb{Y}$. 
% \end{itemize}

The more complicated operations, such as truncation or exponentiation, can be implemented by composing the above basic operations~\cite{10.1007/978-3-642-15317-4_13, div2mp}. Besides combining basic operations, some MPL frameworks~\cite{mohassel2018aby3, aby,wagh2020falcon} provide special optimizations for complicated operations. For example, Mohassel et al.~\cite{mohassel2018aby3} design an efficient comparison protocol in \texttt{ABY3}. In this case, the complicated operations with the special optimizations can be viewed as basic operations of the corresponding MPL framework.
 \begin{figure}[htbp]
    \centering
    \includegraphics[width=0.33\textwidth]{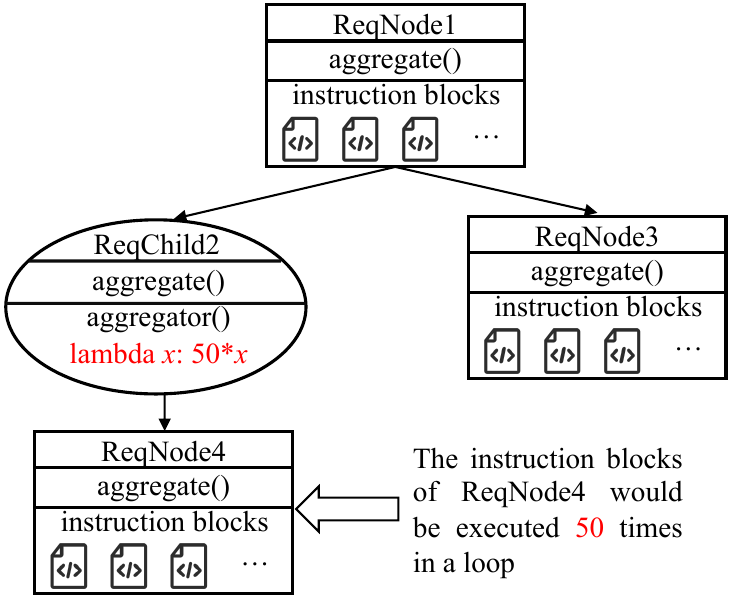}
    \caption{An example of the \mpspdz compiler's block tree. Rectangles and ovals represent ReqNodes and ReqChilds.}
    \label{fig:overview_mpspdz}
\end{figure}

\subsection{\mpspdz Compiler}~\label{subsec:mpspdz_compiler}
We build upon the \mpspdz compiler to construct \hawkeye. The \mpspdz compiler translates \textit{mpc} programs written in Python into a sequence of instructions that represent basic operations of MPL frameworks and store instructions in instruction blocks that contain instructions without branching. Through organizing instruction blocks generated from an \textit{mpc} program as a block tree, the \mpspdz compiler can output the number of basic operations required by the \textit{mpc} program.

As is shown in Figure~\ref{fig:overview_mpspdz}, the block tree of the \mpspdz compiler contains two types of nodes: ReqNode and ReqChild. ReqNode stores a list of instruction blocks. ReqChild records the control flow information. The root node of a block tree is a ReqNode. Concretely, a ReqNode instance contains an aggregate function, a list of instruction blocks, and a list of children, which could be ReqNode or ReqChild. The aggregate function of a ReqNode instance is used to count the number of basic operations in its instruction blocks and children. A ReqChild instance contains an aggregator function, an aggregate function, and a list of children, which must be ReqNode. The aggregator function of a ReqChild instance is an anonymous function that contains the control flow information. For example, in Figure~\ref{fig:overview_mpspdz}, the aggregator function of ReqChild2 is an anonymous function that multiplies input data (\wqruanother{i.e.,} the operation statistics from its children) fifty times because the instructions in ReqNode4 represent a loop body in a loop whose size is $50$. The aggregate function of a ReqChild is used to sum the operation statistics outputted by its aggregator function. After the \mpspdz compiler generates the block tree from an \textit{mpc} program, it recursively calls the aggregate functions of nodes in the block tree to compute the number of operations required by the \textit{mpc} program. 

\subsection{Automatic Differentiation}~\label{subsec:pre_autograd}
Autograd automatically computes the derivative (gradients) of parameters to simplify model construction. It automatically computes the partial derivative of a function by expressing the function as a sequence of basic operators. The partial derivatives of output to the intermediate values and inputs are computed by applying the chain rule to these basic operators. Therefore, with autograd technology, model designers only need to define the forward process of models, and the gradients of model parameters can be automatically computed in the backward process. Following \texttt{PyTorch}, we design and implement the Autograd library of \hawkeye based on operator overloading-based autograd technology. 

\begin{figure}[htbp]
    \centering
    \includegraphics[width=0.35\textwidth]{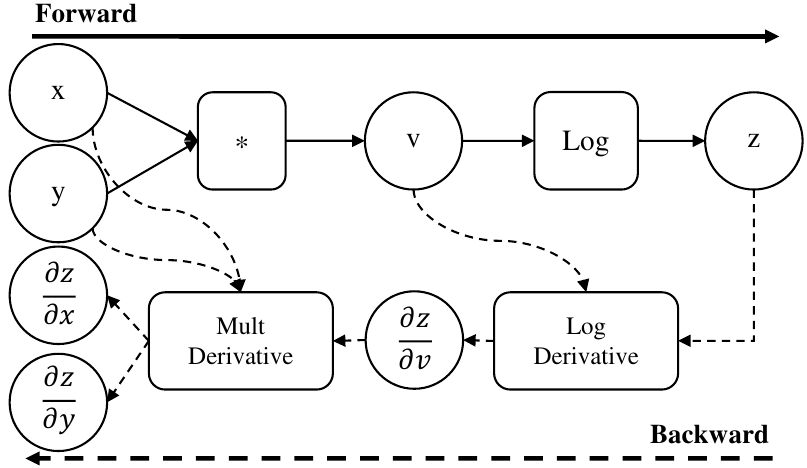}
    \caption{An example of the operator overloading-based autograd. The solid lines represent the forward process, and the dashed lines represent the backward process.}
    \label{fig:augograd}
\end{figure}

\begin{figure*}[htbp]
    \centering
    \includegraphics[width=0.98\textwidth]{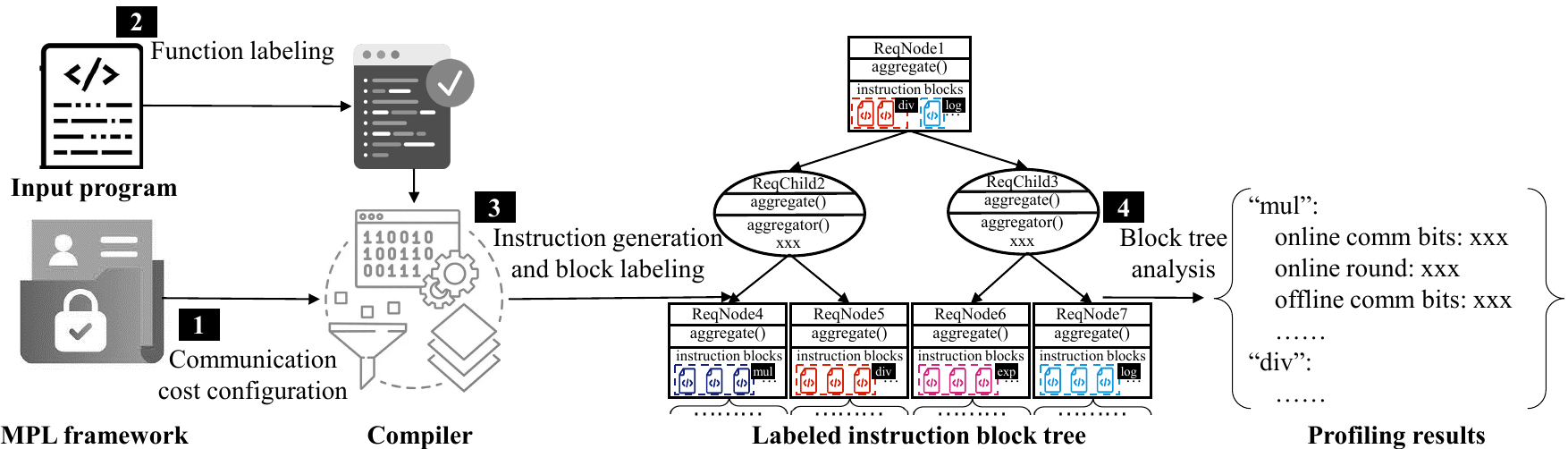}
    \caption{The workflow of our proposed static communication cost profiling method. }
    \label{fig:workflow}
\end{figure*}
The main idea of operator overloading-based autograd technology is to overload basic operators so that each basic operator contains a forward function and a derivative computation function. When model designers use one overloaded basic operator in the forward phase, the basic operator is recorded in an operator list for derivative computations. After the forward process is finished and the backward process starts, the derivative computation function of each overloaded operator is sequentially called from the end of the operator list to compute the derivative of outputs to the intermediate values and inputs. In this way, model designers can use overloaded basic operators to construct complex functions and automatically compute the gradients of the target function. For example, as is shown in Figure~\ref{fig:augograd}, in the forward phase, to compute the output $z$, model designers first obtain $v$ by multiplying $x$ and $y$. Then, model designers obtain $z$ by computing the logarithm of $v$. The above two operators are recorded in a list during the forward phase. In the backward phase, the derivative of $\frac{\partial z}{\partial v}$ is computed by calling the derivative computation function of the logarithm operator with $z$ and $v$ as inputs. Then, $\frac{\partial z}{\partial x}$ and $\frac{\partial z}{\partial y}$ are computed by calling the derivative computation function of the multiplication operator with $\frac{\partial z}{\partial v}$, $x$ and $y$ as inputs.

%% file: tex/3.Cost_analysis.tex
\section{Static Communication Cost Profiling Method}~\label{sec:analysis} 
% \subsection{Overview}~\label{subsec:profiling_overview}
As is shown in Figure~\ref{fig:workflow}, the workflow of our proposed static communication cost profiling method compromises four steps: (1) Model designers choose an MPL framework to perform MPL tasks and configure the communication cost of the MPL framework with the communication cost configuration interface of \hawkeye. (2) Given an \textit{mpc} program, model designers label the functions in the \textit{mpc} program they want to profile with the function labeling interface of \hawkeye. (3) \hawkeye generates the instruction block tree of the labeled \textit{mpc} program and assigns each generated instruction block a label that corresponds to the labeled function based on a prefix structure. (4) \hawkeye analyzes the labeled block tree and outputs the profiling results. Note that we have labeled forward and derivative computation functions of operators that require communication in the Autograd library of \hawkeye. Thus, model designers can directly apply \hawkeye to profile the model communication cost in \texttt{PyTorch} without manually labeling the model components.

\subsection{Communication Cost Configuration and Function Labeling Interfaces }~\label{subsec:label_interface}
\noindent \textbf{Communication cost configuration interface.} 
The communication cost of one basic operation typically depends on the following five parameters: bit length $k$, statistical security parameter $\kappa_s$, computational security parameter $\kappa$, bit length $f$ of a fixed-point number's fractional part, the number of parties $m$. The five parameters should be set according to practical requirements. As shown in Listing~\ref{list:cost}, in \hawkeye, the communication cost of instruction is expressed as an anonymous function that takes $k$, $\kappa_s$, $\kappa$, $f$, and $m$ as inputs and outputs a tuple that contains online communication size, online communication round, offline communication size, offline communication round. \wqruan{Some operations could have extra parameters. For example, the communication cost function of secure multiplication has three extra parameters: the degree of HE polynomial $\mathit{deg}$, HE prime coefficients modulus $\mathit{mod}$, and $\mathit{size}$ that indicates the number of multiplication operations executed in parallel. We describe them in detail in Appendix~\ref{appendix:protocol_config}.}
% For the basic instruction corresponding to Matrix\_Multiplication, its communication cost also depends on the shape of input matrices. Therefore, the communication cost function of Matrix\_Multiplication has three extra parameters: $p$, $q$, and $r$, representing the row number and the column number of the first input matrix and the column number of the second input matrix, respectively. 
% Note that the communication cost of basic operations is typically analyzed in the original paper of the corresponding MPL framework. Therefore, model designers can configure the communication cost for basic operations of one MPL framework according to the original paper of the MPL framework.
% The communication cost configuration process should be much easier than implementing MPC-friendly models for MPL frameworks. 
\begin{lstlisting}[mathescape,xleftmargin=2em,framexleftmargin=2em, caption = {Communication cost configuration for part of basic operations of $\texttt{ABY3}$~\cite{mohassel2018aby3} framework. One basic operation corresponds to one basic instruction in cost\_function\_dict}, columns=fullflexible, label = {list:cost}]
class ABY3(Cost):
  cost_func_dict = {
    "share": lambda $k$, $\kappa_s$, $\kappa$, $f$, $m$: (3*$k$, 1, 0, 0),
    "reveal": lambda $k$, $\kappa_s$, $\kappa$, $f$, $m$: (3*$k$, 1, 0, 0)
    "muls": lambda $k$, $\kappa_s$, $\kappa$, $f$, $m$, $\mathit{size}$, $\mathit{deg}$, $\mathit{mod}$: 
    (3*$k$*$\mathit{size}$, 1, 0, 0)
\end{lstlisting}

In addition to the above basic operations, complicated operations, \wqruanother{e.g.,} truncation or exponentiation, are implemented by composing basic operations by default in \hawkeye. However, if model designers use an MPL framework with special optimizations for some complicated operations (\wqruan{e.g., the comparison operation based on mixed protocols}), they can configure the communication cost of the complicated operations. \wqruan{For example, many mixed-protocol MPL frameworks~\cite{aby, mohassel2018aby3, mishra2020delphi, crypten2020} implement the non-linear operations by converting arithmetic shares to boolean shares or Yao shares. To handle this type of MPL framework, model designers can specify the protocol assignments for the non-linear operations and configure their communication cost, which includes the communication cost of share conversion and boolean circuit evaluation. We further discuss how to adaptively assign protocols for non-linear operations of mixed-protocol frameworks in Section~\ref{sec:dis}.}

\wqruan{The communication cost of MPL frameworks can be configured through the following three methods: (1) Analyzing the communication costs of basic operations according to the description of the original paper. (2) Running the basic operations with open-source codes of MPL frameworks to test the communication costs of basic operations. (3) Asking the framework developers on open-source platforms to further align the communication cost configuration with the concrete MPL framework implementations. The above three methods can be solely used or combined to obtain accurate communication cost configuration. In addition, once the communication cost for one MPL framework is configured, the configuration can be shared with other users through open-source platforms.}

\wqruan{In \hawkeye, to assist model designers in profiling model communication cost on multiple MPL frameworks, we have configured the communication cost of ten MPL frameworks (e.g., \texttt{ABY3}~\cite{mohassel2018aby3}). Thus, model designers can directly apply \hawkeye to profile model communication costs on the ten MPL frameworks. We show the communication cost configuration of the ten MPL frameworks in Appendix~\ref{appendix:protocol_config}.}

\wqruan{Besides, when setting the value of $k$ and $f$, model designers should be careful about the impact of the truncation method. In particular, local truncation, i.e., the truncation method used by \texttt{SecureML}~\cite{DBLP:conf/sp/MohasselZ17} and \texttt{ABY3}~\cite{mohassel2018aby3}, requires additional bits (slack bits) to maintain the desired truncation failure rate. Otherwise, the truncation results would suffer the sign bit flip error with a relatively high probability. Therefore, if using \texttt{ABY3/SecureML}-style truncation, model designers should preserve slack bits when setting the value of $k$ and $f$.}

\begin{lstlisting}[mathescape,xleftmargin=2em,framexleftmargin=2em, caption = {An example program whose functions are labeled using the function labeling interface of \hawkeye.},columns=fullflexible, label = {list:function_labeling}]
@buildingblock("mul") #label mul function
def mul(a, b):
    return a*b   
@buildingblock("test") #label test function
def test(a, b):
    c = mul(a,b)
    n = reveal(c)
    return n
x = sint(1), y = sint(2)
test(x, y)
\end{lstlisting}

\noindent \textbf{Function labeling interface.} \hawkeye provides a function decorator (\wqruanother{i.e.,} @buildingblock) for model designers to label the functions. By adding @buildingblock(``\textit{func1}'') before the declaration of a function, model designers can assign the label ``\textit{func1}'' to the function. For example, as is shown in Listing~\ref{list:function_labeling}, we assign labels ``mul'' and ``test'' to \textit{mul} and \textit{test} functions with the function decorator. In the main body of the program, we call the \textit{test} function with two secret shared integers as inputs. When the bit length is 64 and model designers apply \hawkeye to profile the program shown in Listing~\ref{list:function_labeling} with the communication cost configuration shown in Listing~\ref{list:cost}, the profiling result outputted by \hawkeye would be \{ ``initial-test'': (192, 1, 0, 0), ``initial-test-mul'': (192, 1, 0, 0) \}. The four numbers in tuples of the above dictionary represent online communication bits, online communication round numbers, offline communication bits, and offline communication round numbers, respectively. 
In the above dictionary, the communication cost of \textit{test} function is the sum of all items whose labels contain ``test'', and the same goes for \textit{mul} function. The format of the profiling result originated from our proposed blocking labeling methods, which are described in Section~\ref{subsec:block_label_analysis}.

\subsection{Block Labeling and Tree Analysis Method}~\label{subsec:block_label_analysis}
After model designers label the functions in an \textit{mpc} program with the function labeling interface of \hawkeye, \hawkeye generates and labels instruction blocks to obtain the labeled block tree. Then, \hawkeye analyzes the labeled block tree to obtain the communication cost profiling results. 

% \smallskip
\noindent\textbf{Block labeling method.} To label instruction blocks for accurate communication cost profiling, we need to resolve the following issue: one labeled function could call another labeled function. In this situation, directly using the label of the called function to label generated instruction blocks would cause the information of the function call chain to be lost.  To resolve the issue, we propose a prefix structure to record the information of the function calling relations. Specifically, we maintain a global label. When compiling a labeled function, we append the label of the function to the global label and use it to label the generated instruction blocks. After generating instructions corresponding to the function, we remove its label from the end of the global label. In this way, when a labeled function calls another labeled function, the information of the calling function would be recorded in the global label. 
% Meanwhile, the communication cost of one labeled function would be the sum of all items in profiling results whose keys contain its label.

As shown in Algorithm~\ref{alg:block_label}, \hawkeye labels instruction blocks in a recursive manner. We first define two functions, Compile\_S\_List and Compile\_Func. Compile\_S\_List (Line 4-9) receives a statement list S\_List, the block tree $\mathcal{T}$, and the global label g\_label as inputs. It compiles the statements of S\_List as instruction blocks and labels generated instruction blocks with g\_label. The inputs of Compile\_Func (Line 10-23) contain a function $func$,  $\mathcal{T}$, g\_label, and S\_List. During the instruction generation process, Compile\_Func enumerates statements of $func$. If one statement $s$ does not call a labeled function $f_{sub}$, Compile\_Func appends $s$ into S\_List (Line 19). Otherwise, Compile\_Func first compiles the statements stored in the S\_List (Line 14). Then, Compile\_Func appends the label of $f_{sub}$ to the end of g\_label (Line 15). Without loss of generality, we assume that $s$ is a call to $f_{sub}$ and does not perform other computations. After that, Compile\_Func calls Compile\_Func with $f_{sub}$, $\mathcal{T}$, and g\_label as inputs and then removes the label of $f_{sub}$ from the end of g\_label (Line 16-17). Compile\_Func clears S\_List by calling Compile\_S\_List (Line 22). Finally, after initializing g\_label, $\mathcal{T}$, and S\_List, block labeling process is completed by calling Compile\_Func with $p$, $\mathcal{T}$, g\_label and S\_List as inputs (Line 2). Note that in Algorithm~\ref{alg:block_label}, we view the input program $p$ as a function that contains a sequence of statements.

% \smallskip
\noindent\textbf{Block tree analysis method.} We profile the communication cost of an \textit{mpc} program by analyzing its corresponding labeled block tree $\mathcal{T}$ outputted by Algorithm~\ref{alg:block_label}. 
As is shown in Algorithm~\ref{alg:block_analysisl}, the aggregate function of a ReqNode instance receives a dictionary $d_1$ as the input and updates $d_1$ with the communication cost profiling results of the block tree whose root node is the ReqNode instance. For a ReqNode instance, its aggregate function first calculates the communication cost of each instruction block in its instruction block list (Line 5-12). Given an instruction block $b$, the aggregate function enumerates each instruction $i$ in $b$ and computes the communication cost of $i$ by calling its communication cost function defined in the communication cost configuration dictionary (Line 8). Then, the aggregate function adds the cost function output to $d_1$ according to the label of $b$ (Line 9). After calculating the communication cost of the ReqNode instance itself, its aggregate function calls the aggregate function of its children with $d_1$ as the input (Line 13-15).

\begin{algorithm}[htbp]
\small
\caption{Block labeling algorithm that generates and labels instruction blocks. }
\label{alg:block_label}
 \begin{algorithmic}[1]
    \REQUIRE An \textit{mpc} program $p$ that represents as a sequence of statements. The functionality \textit{Compile\_Statements} receives a statement list and the block tree $\mathcal{T}$ as input and compiles these statements into instruction blocks that are inserted into $\mathcal{T}$. The \mpspdz compiler provides it.
    \ENSURE The labeled block tree $\mathcal{T}$.
    \STATE \textbf{Initialization}: initialize the global label as g\_label = `initial', the block tree $\mathcal{T}$ as an empty ReqNode instance, a statement list S\_List as []. 
    \STATE Compile\_Func(p, $\mathcal{T}$, g\_label, S\_List)
    \RETURN $\mathcal{T}$
    \STATE\textbf{Function Compile\_S\_List(}S\_List, $\mathcal{T}$, g\_label\textbf{):}
    \begin{ALC@g}
    \STATE blocks = \textit{Compile\_Statements}(S\_List, $\mathcal{T}$)
    \FOR{block $b$ in blocks}
    \STATE b.label = g\_label
    \ENDFOR
    \STATE S\_List = []
    \end{ALC@g}
    \STATE \textbf{End Function}
    \STATE\textbf{Function Compile\_Func}(\textit{func}, $\mathcal{T}$, g\_label, S\_List\textbf{):}
    \begin{ALC@g}
    \FOR{statement $s$ in \textit{func}}
    \IF{$s$ calls a labeled function $f_{sub}$}
    \STATE Compile\_S\_List(S\_List, $\mathcal{T}$, g\_label)
    \STATE g\_label = g\_label + `-$f_{sub}$.label'
    \STATE Compile\_Func($f_{sub}$, $\mathcal{T}$, g\_label, S\_List)
    \STATE g\_label = g\_label - `-$f_{sub}$.label'
    \ELSE
    \STATE S\_List.append($s$)
    \ENDIF
    \ENDFOR
    \STATE Compile\_S\_List(S\_List, $\mathcal{T}$, g\_label)
    \end{ALC@g}
    \STATE\textbf{End Function}
\end{algorithmic}
\end{algorithm}

The aggregate function of a ReqChild instance receives a dictionary $d_1$ as the input. It processes the communication cost profiling results outputted by its children with its aggregator function and updates $d_1$ with the process results.  For a ReqChild instance, its aggregate function first initializes an empty dictionary \textit{tmp\_d} (Line 18). Then, the aggregate function of the ReqChild instance calls the aggregate functions of its children one by one with \textit{tmp\_d} as the input (Line 19-21). After that, for each pair in \textit{tmp\_d}, the aggregate function processes the item of the pair with its aggregator function and adds the result to $d_1$ according to the label of the pair (Line 22-24). Finally, the block tree analysis process is completed by initializing an empty dictionary $d$ and calling the aggregate function of \textit{root} with $d$ as input, where \textit{root} is the root ReqNode of the labeled block tree $\mathcal{T}$ (Line 1-3).
\begin{algorithm}[htbp]
\small
\caption{Block tree analysis algorithm that analyzes the labeled block tree to profile the communication cost of an \textit{mpc} program. The addition between two tuples outputs a new tuple whose elements are the sum of two input tuples' corresponding elements}
\label{alg:block_analysisl}
 \begin{algorithmic}[1]
    \REQUIRE The root node \textit{root} of the labeled block tree $\mathcal{T}$ and a \textit{cost\_func\_dict} stores communication cost functions of basic operations of the used MPL framework.
    \ENSURE A dictionary $d$ that stores the profiling result.
    \STATE \textbf{Initialization:}  Initialize an empty dictionary $d$
    \STATE \textit{root}.aggregate($d$)
    \RETURN $d$
    \STATE\textbf{Function ReqNode: aggregate(}self, $d_1$\textbf{):}
    \begin{ALC@g}
    \FOR{block $b$ in self.blocks}
    \FOR{instructions $i$ in $b$}
        \IF{$i$ requires communication}
          \STATE  i\_cost = \textit{cost\_func\_dict}[$i$.name]($i$.args) 
            \STATE   $d_1$[$b$.label] = $d_1$[$b$.label] + i\_cost
        \ENDIF
    \ENDFOR
    \ENDFOR
    \FOR{child $c$ in self.children}
      \STATE  $c$.aggregate($d_1$)
    \ENDFOR
    \end{ALC@g}
    \STATE \textbf{End Function}
    \STATE\textbf{Function ReqChild: aggregate(}self, $d_1$\textbf{):}
    \begin{ALC@g}
    \STATE Initialize an empty dictionary \textit{tmp\_d}
    \FOR{node $n$ in self.children}
      \STATE  $n$.aggregate(\textit{tmp\_d})
    \ENDFOR
    \FOR{label, item in \textit{tmp\_d}}
            \STATE   $d_1$[label] =  $d_1$[label] + self.aggregator(item)
    \ENDFOR
    \end{ALC@g} 
    \STATE \textbf{End Function}
\end{algorithmic}
\end{algorithm}

\noindent\textbf{Accuracy analysis.} Because of the security requirements of MPL, the communication cost profiling results outputted by the static communication cost profiling method are consistent with those obtained by dynamic profiling. Concretely, to rigorously guarantee the data security of MPL frameworks, the program execution flow of MPL frameworks must be input-independent~\cite{10.1145/28395.28416}. Therefore, the execution flow of each labeled function, which can be viewed as a sequence of instructions, must remain unchanged as input data changes. As a result, given an \textit{mpc} program that contains labeled functions, as long as model designers correctly configure the communication cost for basic operations of MPL frameworks, \hawkeye can accurately profile the communication cost of the \textit{mpc} program.

%% file: tex/4.Autograd.tex
\section{Autograd Library of \hawkeye}~\label{sec:autograd}
% \vspace{-6mm}
\subsection{Architecture}~\label{subsec:autograd_architecture}
\begin{figure}[htbp]
    \centering
    \includegraphics[width=0.42\textwidth]{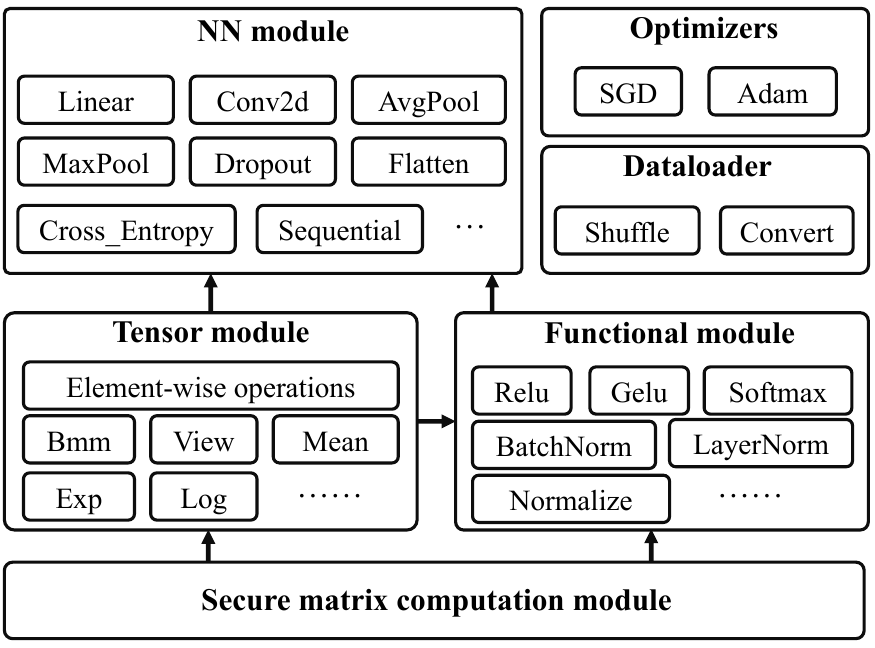}
    \caption{The architecture of the \hawkeye Autograd library.  }
    \label{fig:overview-autograd}
\end{figure}

As is shown in Figure~\ref{fig:overview-autograd}, the Autograd library of \hawkeye contains six modules: Secure matrix computation module, Tensor module, Functional module, NN module, Optimizers, and Dataloader. The Secure matrix computation module is the basis of the other five modules and would not be exposed to model designers. The interfaces of the other five modules are fully consistent with those of \texttt{PyTorch}. We then describe the above six modules as follows.

    \noindent \textbf{(1) Secure matrix computation module.} This module contains secure matrix computation functions that are necessary for MPL. These functions are either our supplemented computation functions (\wqruanother{e.g.,} permutation) or provided by the original \mpspdz compiler (\wqruanother{e.g.,} matrix multiplication). The module is the basis of the \hawkeye Autograd library. All of the other five modules are built on this module.
    
    \noindent \textbf{(2) Tensor module.} This module contains tensor computation operators, such as element-wise operators and batch matrix-matrix products. As is mentioned in Section~\ref{subsec:pre_autograd}, each tensor computation operator of the Tensor module is overloaded to contain a forward function and a derivative computation function, such that the gradients of parameters would be automatically derived. We label the forward and derivative computation functions of each operator of this module to support model communication cost profiling.
    
    \noindent \textbf{(3) Functional module.} This module contains activation functions (\wqruanother{e.g.,} Relu) and functions that depend on model parameters (\wqruanother{e.g.,} BatchNorm). Like the Tensor module, each operator of this module contains a labeled forward function and a labeled derivative computation function.
    
    \noindent  \textbf{(4) NN module.} This module contains operators (\wqruanother{e.g.,} Conv2d) and containers (\wqruanother{e.g.,} Sequential) for model construction. The operators of the NN module are implemented using functions from the Tensor module and the Functional module. Like \texttt{PyTorch}, model designers can also define new operators based on operators of the NN module, the Tensor module, and the Functional module.
    
    \noindent  \textbf{(5) Optimizers.} Optimizers contain popular optimizers in machine learning, \wqruanother{e.g.,} stochastic gradient descent (SGD) and Adam~\cite{adam}. Their step functions that require communication are also labeled to profile the secure model training processes.
    
    \noindent \textbf{(6) Dataloader.} Dataloader shuffles the input data and converts the input data into tensors.

\subsection{Optimization of Broadcast Operator}
\begin{figure}[htbp]
    \centering
    \subfloat[Forward phase 
    \label{fig:broadcast_forward}]
    {
        \includegraphics[width=1.9cm]{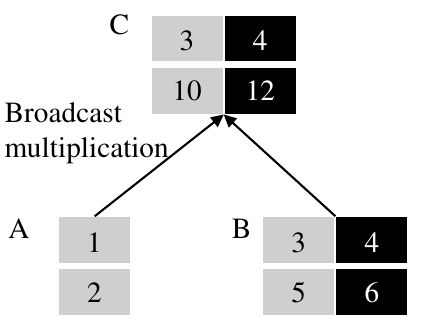}
    }\hfill
    \subfloat[Strawman backward phase
    \label{fig:broadcast_strawmanbackward}]
    {
        \includegraphics[width=2.2cm]{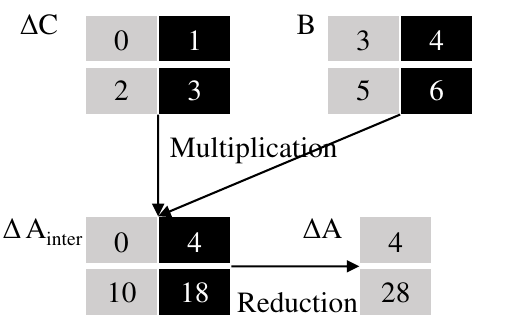}
    }\hfill
    \subfloat[Optimized backward phase \label{fig:broadcast_optimizaedbackward}]
    {\includegraphics[width=2.9cm]{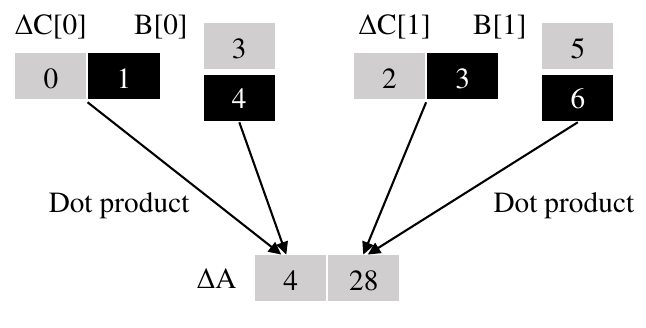}}
    \caption{The strawman implementation of broadcast multiplication and optimized broadcast multiplication.}
    \label{fig:broadcast}
\end{figure}

\noindent\textbf{Strawman implementation of the broadcast operator.} We first describe the strawman implementation of the broadcast operator and show that the implementation would bring extra communication overhead. Taking broadcast multiplication as an example, as shown in Figure~\ref{fig:broadcast_forward}, given two inputs $A$, $B$ with shapes (1, 2) and (2, 2), broadcast multiplication performs element-wise multiplication between each column of $B$ and $A$ to output $C$. During the backward phase, to compute the derivative of $A$, as is shown in Figure~\ref{fig:broadcast_strawmanbackward}, a straightforward method is first performing element-wise multiplication between $\Delta C$ and $B$ to obtain an intermediate result $\Delta A_{inter}$. Then, we can sum each column of $\Delta A_{inter}$ to reduce $\Delta A_{inter}$ as $\Delta A$. The main drawback of the strawman implementation is that it produces the intermediate result with a larger shape than the final result, which causes extra communication overhead. The main reason is that the communication costs of some MPL frameworks, \wqruanother{e.g.,} \texttt{ABY3}, \wqruanother{only depend on the size of the computation results and are independent of the size of input data and intermediate results.} Therefore, the intermediate result $\Delta A_{inter}$ whose size is two times the size of the final result $\Delta A$ would bring significantly more communication overhead, causing bias in the static profiling results from \hawkeye.

\noindent\textbf{Optimized broadcast operator.} To address the drawback of the strawman implementation, we fuse the broadcast computation and intermediate result reduction as a parallel vector computation to compute the final result directly. Concretely, as shown in Figure~\ref{fig:broadcast_optimizaedbackward}, the backward process of the optimized broadcast operator is completed by performing the dot product between each row of $\Delta C$ and $B$ in parallel. In this way, we can avoid the intermediate result with a large size, thus avoiding extra communication overhead. For more complex broadcast cases, we can transform the input data into the two-dimensional case shown in Figure~\ref{fig:broadcast} through reshaping and permuting. For example, for two inputs with shapes (5, 3, 2) and (3, 2), we can reshape them as (5, 6) and (1, 6) and finish the broadcast computation in the optimized way.

\subsection{Integration of Model Communication Cost Profiling Method}~\label{subsec:autograd_profiling}
We then describe how the Autograd library of \hawkeye integrates the static model communication cost profiling method introduced in Section~\ref{sec:analysis}. We first discuss two requirements that the Autograd library of \hawkeye should meet to enable model designers to profile the communication cost of models in \texttt{PyTorch}. Firstly, the communication cost of the model forward and backward processes should be profiled automatically. Manually inserting test instruments in secure model training or inference codes could be time-consuming for model designers. Meanwhile, even though a few MPL frameworks (\wqruanother{e.g.,} \texttt{CrypTen}~\cite{crypten2020}) provide user-friendly model construction interfaces to assist model designers in building a secure model training or inference process, they hide the derivative computation functions of operators. Thus, model designers have to modify the source codes of these MPL frameworks to profile the communication cost of model backward processes. The code modification process would require burdensome human efforts. Therefore, the Autograd library of \hawkeye should automatically profile the communication cost of the model forward and backward processes.  

\begin{lstlisting}[escapechar=!,mathescape,xleftmargin=2em,framexleftmargin=2em, caption = {The labeled exp function in the Tensor module of the Autograd library of \hawkeye. The newly added labeling codes are highlighted on a green background and labeled with a plus sign at the beginning of the line.},columns=fullflexible, label = {list:autograd_labeling}]
!\colorbox[RGB]{230,255,236}{$+$ @buildingblock("exp-forward")}!
def exp(self):
!\colorbox[RGB]{230,255,236}{$+ \;$ @buildingblock("exp-backward")}!
    def propagate(dl_doutputs, operation):
        ...... 
        #The derivative computation process of exp
        return dl_dinputs
    ..... #The forward process of exp
    return result
\end{lstlisting}

Secondly, the granularity of model communication cost profiling can be adjusted without manually inserting test instruments. To simplify the model construction process, model designers usually construct models in a hierarchical manner. For example, one Densenet-121 model~\cite{Huang_2017_CVPR} contains multiple Bottlenecklayers and TransitionLayers. These two types of sub-modules both contain Conv2d operators, pooling operators, Relu functions, etc. To comprehensively profile the model communication cost, model designers usually need to test the communication cost of operators at different granularity. 

\begin{lstlisting}[escapechar=!,mathescape,xleftmargin=2em,framexleftmargin=2em, caption = {An example of labeling the forward functions of sub-modules to adjust the granularity of model communication cost profiling. The newly added labeling codes are highlighted on a green background and labeled with a plus sign at the beginning of the line.},columns=fullflexible, label = {list:component_labeling}]
class TransitionLayer(nn.Module):
    def __init__(self, c_in, c_out):
        super().__init__()
        self.layers = nn.Sequential(
            #label: transitionlayer-batchnorm
            nn.BatchNorm2d(num_features=c_in),
            #label: transitionlayer-relu
            nn.ReLU(),
            #label: transitionlayer-conv2d
            nn.Conv2d(in_channels=c_in, 
                    out_channels=c_out, kernel_size=1),
            #label: transitionlayer-avgpool2d
            nn.AvgPool2d(kernel_size=2))
!\colorbox[RGB]{230,255,236}{$+ \;$  @buildblock("transitionlayer")}! #labeling transitionlayer
    def forward(self, x):
        out = self.layers(x)
        return out
\end{lstlisting}

We meet the first requirement by labeling the forward and derivative computation functions of each operator and function that requires communication. For example, as is shown in Listing~\ref{list:autograd_labeling}, the exp function in the Tensor module contains a propagate function that defines the derivative computation process of the exponentiation operator, and the rest part defines its forward process. To profile the communication cost of the forward and backward processes of the exponentiation operator, we label the exp function and its propagate function with the function labeling interface described in Section~\ref{subsec:label_interface}. When model designers use the exponentiation operator to construct models and profile them with \hawkeye, the communication cost of the exponentiation operator can be obtained by summing the items whose keys contain ``\textit{exp-forward}'' or ``\textit{exp-backward}'' in the profiling results. Furthermore, the communication cost of the exponentiation operator forward process can be separately computed by summing the items whose labels contain ``\textit{exp-forward}''. We apply the above labeling process to each operator and function that requires communication in the Autograd library of \hawkeye. As a result, \hawkeye can automatically profile the communication cost of the model forward and backward processes.

For the second requirement, \hawkeye meets it based on the prefix structure of our proposed static communication cost profiling method. Concretely, model designers can adjust the granularity of model communication cost profiling by labeling the forward functions of sub-modules with the function labeling interface described in Section~\ref{subsec:label_interface}. Taking the Densenet-121 model as an example, as mentioned above, two sub-modules of the Densenet-121 model, \wqruanother{i.e.,} Bottlenecklayers and TransitionLayers, both contain basic operators (\wqruanother{e.g.,} Conv2d) of the \hawkeye Autograd library. If model designers do not label forward functions of these two sub-modules, \hawkeye would output the total communication cost of basic operators without distinguishing the sub-modules these basic operators belong to. In contrast, as is shown in Listing~\ref{list:component_labeling}, if model designers label the forward function of the TransitionLayer with the label ``\textit{transitionlayer}'', the communication cost of basic operators in TransitionLayer would be outputted separately with the prefix ``\textit{transitionlayer}''. Because composing sub-modules is a common way to construct complex models, labeling the forward functions of sub-modules should be a promising way for model designers to adjust the granularity of model communication cost profiling.

%% file: tex/5.Experiments.tex
\begin{table*}[ht]
\centering
\caption{\wqruan{The online communication size profiling results of four secure CNN model inference processes outputted by \texttt{CryptFlow2}~\cite{rathee2020cryptflow2} and \hawkeye.} We report the proportion of each operator's online communication size to the total online communication size and the online communication size of each operator. Following Ganesan et al.~\cite{ganesan2022efficient}, we list the online communication size across linear and non-linear operators. }
\scalebox{0.7}{
\begin{tabular}{c|c|cccc}
\hline
Model                          & Framework                                                                              & \multicolumn{3}{c}{\% (GB) of linear operators}                                                                                                                                                                                                                                                      & \% (GB) of non-linear operators                                                       \\ \hline
\multicolumn{1}{l|}{}         &                                                                                        & \multicolumn{1}{c}{\% (GB) of Conv2d}                                                               & \multicolumn{1}{c}{\% (GB) of other linear operators}                                                       & \% (GB) of all                                                                     & \multicolumn{1}{l}{}                                                        \\ \hline
\multirow{3}{*}{DenseNet-121} & \texttt{CrypTFlow2}~\cite{rathee2020cryptflow2} & \multicolumn{1}{c}{\wqruan{94.14\% (646.05GB)}}                                                              & \multicolumn{1}{c}{\wqruan{3.60\% (24.72GB)}}                                                              & \wqruan{97.74\% (670.78GB)}                                                              & \wqruan{2.26\% (15.52GB)}                    \\ \cline{2-6} 
                 & \hawkeye                                                                & \multicolumn{1}{c}{\begin{tabular}[c]{@{}c@{}}\wqruan{93.92\% (657.06GB)}\\ \wqruan{-0.22\% (+11.01GB)}\end{tabular}} & \multicolumn{1}{c}{\begin{tabular}[c]{@{}c@{}}\wqruan{3.77\% (26.33GB)}\\ \wqruan{+0.17\% (+1.61GB)}\end{tabular}} & \begin{tabular}[c]{@{}c@{}} \wqruan{97.69\% (683.39GB)}\\ \wqruan{-0.05\% (+12.61GB)}\end{tabular} & \begin{tabular}[c]{@{}c@{}}\wqruan{2.31\% (16.17GB)}\\ \wqruan{+0.05\% (+0.65GB)}\end{tabular}                       \\ \hline
\multirow{3}{*}{ResNet-50}     & \texttt{CrypTFlow2}~\cite{rathee2020cryptflow2} & \multicolumn{1}{c}{\wqruan{96.17\% (851.76GB)}}                                                              & \multicolumn{1}{c}{\wqruan{2.04\% (18.05GB)}}                                                              & \wqruan{98.21\% (869.81GB)}                                                              & \wqruan{1.79\% (15.83GB)}                        \\ \cline{2-6} 
    & \hawkeye                                                                & \multicolumn{1}{c}{\begin{tabular}[c]{@{}c@{}}\wqruan{96.14\% (863.11GB)}\\ \wqruan{-0.03\% (+11.35GB)}\end{tabular}} & \multicolumn{1}{c}{\begin{tabular}[c]{@{}c@{}}\wqruan{2.05\% (18.43GB)}\\ \wqruan{+0.01\% (+0.38GB)}\end{tabular}}  & \begin{tabular}[c]{@{}c@{}}\wqruan{98.19\% (881.54GB)}\\ \wqruan{-0.02\% (+11.73GB)}\end{tabular} & \begin{tabular}[c]{@{}c@{}}\wqruan{1.81\% (16.25GB)}\\ \wqruan{+0.02\% (+0.42GB)}\end{tabular}                                                                 \\ \hline
\multirow{3}{*}{MobileNet-V3}  & \texttt{CrypTFlow2}~\cite{rathee2020cryptflow2} & \multicolumn{1}{c}{\wqruan{75.14\% (73.30GB)}}                                                               & \multicolumn{1}{c}{\wqruan{18.62\% (18.16GB)}}                                                             & \wqruan{93.76\% (91.46GB)}                                                               & \wqruan{6.24\% (6.09GB)}                          \\ \cline{2-6} 
  & \hawkeye                                                                & \multicolumn{1}{c}{\begin{tabular}[c]{@{}c@{}}\wqruan{74.98\% (73.33GB)}\\ \wqruan{-0.16\% (+0.03GB)}\end{tabular}}   & \multicolumn{1}{c}{\begin{tabular}[c]{@{}c@{}}\wqruan{18.57\% (18.16GB)}\\ \wqruan{-0.05\% (+0.00GB)}\end{tabular}} & \begin{tabular}[c]{@{}c@{}}\wqruan{93.55\% (91.49GB)}\\ \wqruan{-0.21\% (+0.03GB)}\end{tabular}   & \begin{tabular}[c]{@{}c@{}}\wqruan{6.45\% \wqruan{(6.31GB)}}\\ \wqruan{+0.21\% (+0.22GB)}\end{tabular}                                                                   \\ \hline
\multirow{3}{*}{ShuffleNet-V2}   & \texttt{CrypTFlow2}~\cite{rathee2020cryptflow2} & \multicolumn{1}{c}{\wqruan{86.18\% (38.42GB)}}                                                               & \multicolumn{1}{c}{\wqruan{9.20\% (4.10GB)}}                                                              & \wqruan{95.38\% (42.52GB)}                                                               & \wqruan{4.62\% (2.06GB)} \\ \cline{2-6} 
& \hawkeye                                                                & \multicolumn{1}{c}{\begin{tabular}[c]{@{}c@{}}\wqruan{86.14\% (38.87GB)}\\ \wqruan{-0.04\% (+0.45GB)}\end{tabular}}   & \multicolumn{1}{c}{\begin{tabular}[c]{@{}c@{}}\wqruan{9.20\% (4.15GB)}\\ \wqruan{+0.00\% (+0.05GB)}\end{tabular}}  & \begin{tabular}[c]{@{}c@{}}\wqruan{95.34\% (43.02GB)}\\ \wqruan{-0.04\% (+0.50GB)}\end{tabular}   & \begin{tabular}[c]{@{}c@{}}\wqruan{4.66\% (2.10GB)}\\ \wqruan{+0.04\% (+0.04GB)}\end{tabular}                                                                                            \\ \hline
\end{tabular}
}
\label{tab:cryptflow2}
\end{table*}

\section{Performance Evaluation}~\label{sec:exp}
\vspace{-6mm}
\subsection{\wqruan{Implementation}} 
We implement \hawkeye by supplementing about 12k lines of code of Python (including test codes) to the \mpspdz compiler. At first, we modify the instruction generation process of the \mpspdz compiler to implement our proposed block labeling method. Then, we modify the aggregate functions of ReqNode and ReqChild in the \mpspdz compiler to implement our proposed block analysis method. 

\wqruan{For the Autograd library of \hawkeye, we implement all of its operators and functions according to the description of \texttt{PyTorch} API document~\footnote{\url{https://pytorch.org/docs/stable/torch.html}}. Concretely, we implement functions of five \texttt{PyTorch}-related modules (Tensor module, Functional module, NN module, Optimizers, and Dataloader) described in Section~\ref{subsec:autograd_architecture} based on the secure fixed-point numbers computation functions provided by the \mpspdz compiler. The above Autograd library modules are integrated into the \mpspdz compiler as one of its standard libraries. Thus, in \hawkeye, when model designers construct models they want to profile, they can directly import the above modules to construct models in \texttt{PyTorch}.}

\begin{figure*}[ht]
    \centering
    \includegraphics[width=0.8\textwidth]{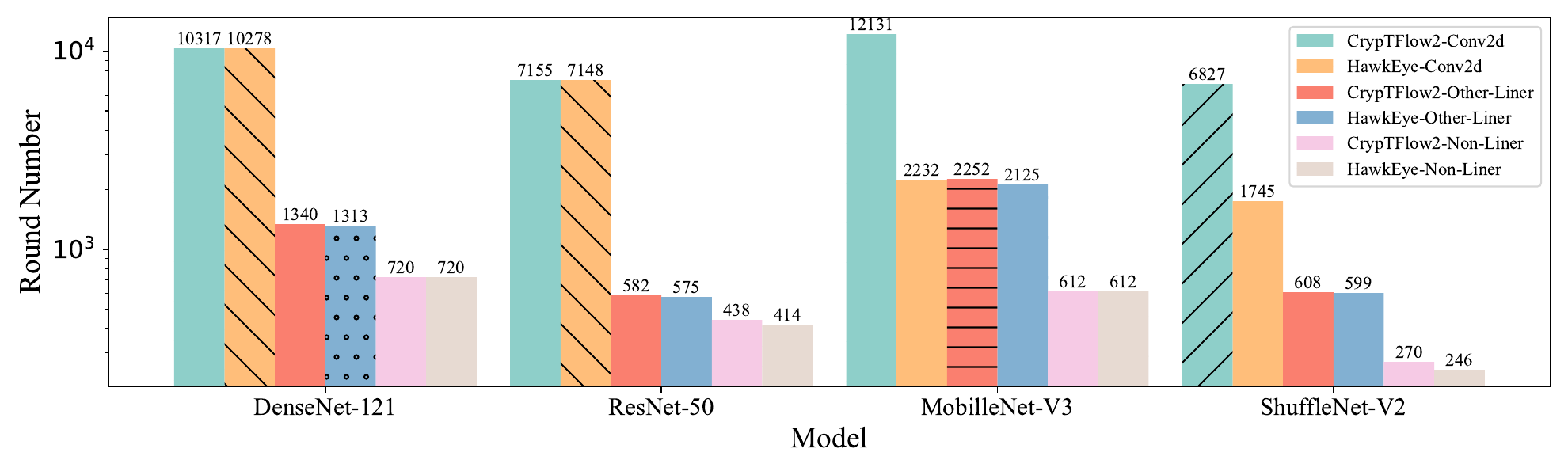}
    \caption{\wqruan{The online communication round profiling results of four secure CNN model inference processes outputted by \texttt{CryptFlow2}~\cite{rathee2020cryptflow2} and \hawkeye.}}
    \label{fig:round_crypTFlow}
\end{figure*}

\subsection{Accuracy of \hawkeye}~\label{subsec:accuracy_garnet}
\begin{table}[ht]
\centering
\caption{\wqruan{The online/offline communication size profiling results of secure BERT$_{\textsc{base}}$ model inference process outputted by \texttt{CrypTen}~\cite{crypten2020} and \hawkeye.} We report the proportion of each operator's communication size to the total communication size and the communication size of each operator. Following Li et al.~\cite{li2023mpcformer}, we list the communication sizes of MatMul, Gelu, and Softmax operators. }
\scalebox{0.7}{
\begin{tabular}{c|c|c|c}
\hline
Operator                 & Framework                                                                  & \% (GB) of online phase           & \% (GB) of offline phase                                                                  \\ \hline
\multirow{3}{*}{MatMul}  & \texttt{CrypTen}~\cite{crypten2020} & \wqruan{4.75\% (3.22GB)}                                         & \wqruan{7.05\% (1.37GB)}                                                                                                                                     \\ \cline{2-4} 
                         & \hawkeye                                                    & \begin{tabular}[c]{@{}c@{}}\wqruan{4.74\% (3.22GB)} \\ \wqruan{-0.01\% (+0.00GB)}\end{tabular} 
                         & \begin{tabular}[c]{@{}c@{}} \wqruan{7.09\% (1.37GB)}\\ \wqruan{+0.04\% (+0.00GB)}\end{tabular}  
                           \\ \hline
\multirow{3}{*}{Gelu}    & \texttt{CrypTen}~\cite{crypten2020} & \wqruan{26.15\% (17.72GB)}                                       & \wqruan{23.87\% (4.64GB)}                                                                                                                                    \\ \cline{2-4} 
                         & \hawkeye                                                    & \begin{tabular}[c]{@{}c@{}}\wqruan{26.49\% (18.00GB)} \\ \wqruan{+0.34\% (+0.28GB)}\end{tabular} 
                         & \begin{tabular}[c]{@{}c@{}} \wqruan{24.12\% (4.64GB)}\\ \wqruan{+0.25\%(+0.00GB)}\end{tabular}        \\ \hline
\multirow{3}{*}{Softmax} & \texttt{CrypTen}~\cite{crypten2020} & \wqruan{69.10\% (46.83GB)}                                            & \wqruan{69.08\% (13.43GB)}                                                                                                                       \\ \cline{2-4} 
                         & \hawkeye                                                    & \begin{tabular}[c]{@{}c@{}}\wqruan{68.77\% (46.73GB)} \\ \wqruan{-0.33\% (-0.10GB)}\end{tabular} 
                         & \begin{tabular}[c]{@{}c@{}} \wqruan{68.79\% (13.23GB)}\\ \wqruan{-0.29\% (-0.20GB)}\end{tabular} \\
                         \hline
\end{tabular}
}
\label{tab:mpcformer}
\end{table}

\begin{figure}[ht]
    \centering
    \includegraphics[width=0.4\textwidth]{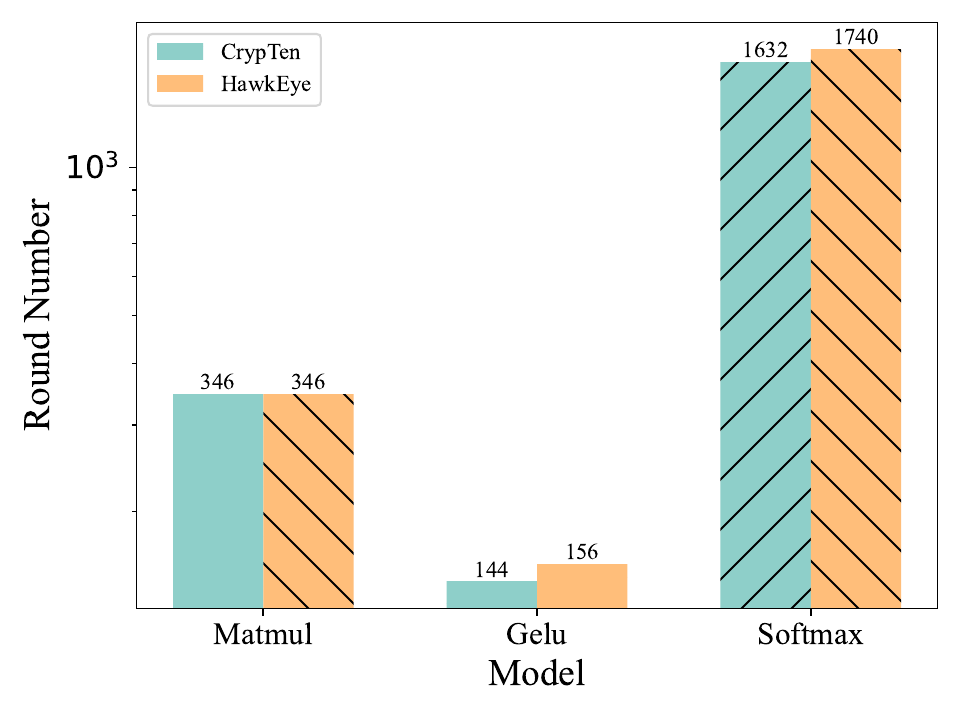}
    \caption{\wqruan{The online communication round profiling results of secure BERT$_{\textsc{base}}$ model inference process outputted by \texttt{CrypTen}~\cite{crypten2020} and \hawkeye.}}
    \label{fig:round_crypTen}
\end{figure}
\noindent\textbf{Experiment Setup.} We verify the accuracy of \hawkeye by comparing the static profiling results outputted by \hawkeye with dynamic profiling results obtained from two MPL frameworks, \wqruanother{i.e.,} \texttt{CypTFlow2}~\cite{rathee2020cryptflow2} and \texttt{CrypTen}~\cite{crypten2020}. We first describe two dynamic profiling processes we compare with.  Firstly, Ganesan et al.~\cite{ganesan2022efficient} dynamically profile secure inference processes of four popular CNN models (\wqruanother{i.e.,} DenseNet-121~\cite{Huang_2017_CVPR}, ResNet-50~\cite{7780459}, MobileNetV3~\cite{howard2019searching}, ShuffleNetV2~\cite{Ma_2018_ECCV}) on \texttt{CrypTFlow2}~\cite{rathee2020cryptflow2}. Secondly, Li et al.~\cite{li2023mpcformer} dynamically profile the secure BERT$_{\textsc{base}}$ model inference process on the two-party backend of \texttt{CrypTen}~\cite{crypten2020}.  We run the open-source codes of the two MPL frameworks~\footnote{The code repository address of \texttt{CrypTFlow2} is \url{https://github.com/mpc-msri/EzPC}. We use the codes at commit \textit{4bae530}.}~\footnote{The code repository address of \texttt{CrypTen} is \url{https://github.com/facebookresearch/CrypTen}. We use \texttt{CrypTen0.4.1}.} to obtain the dynamic profiling results. For dynamic profiling on \texttt{CrypTFlow2}, following Ganesan et al.~\cite{ganesan2022efficient}, we set the bit length as $60$ and the bit length of fixed-point numbers' fractional part as $23$. For dynamic profiling on \texttt{CrypTen}, we set the bit length as $64$ and the bit length of fixed-point numbers' fractional part as $16$. We set the statistical security parameter and computation security parameter of these two MPL frameworks as $40$ and $128$. For the input data of CNN models, we set their sizes as $1 \times 3 \times 224 \times 224$. For the input data of BERT$_{\textsc{base}}$, following Li et al.~\cite{li2023mpcformer}, we set their sizes as $1 \times 512 \times 28996$. Because \texttt{CrypTen} does not support the tanh function, we replace the tanh function with the hardtanh function following Li et al.~\cite{li2023mpcformer}. \wqruan{Note that because \texttt{CrypTFlow2} does not have the offline phase, we only show the offline communication cost profiling results of \texttt{CrypTen}. }
% Besides, because the offline preparation of \texttt{CrypTen} does not depend on circuit structures, it can be finished in the constant round, and we do not consider its offline communication round here.

For static profiling on \hawkeye, we first configure the communication cost of \texttt{CryptoFlow2}~\cite{rathee2020cryptflow2} and \texttt{CrypTen}~\cite{crypten2020}. After that, we implement the models profiled by Ganesan et al.~\cite{ganesan2022efficient} and Li et al.~\cite{li2023mpcformer} in \hawkeye. Finally, we run \hawkeye under the same parameter setting with \texttt{CypTFlow2} and \texttt{CrypTen} to obtain the static profiling results. 

\smallskip
\noindent\textbf{Experimental results.} \wqruan{As is shown in Table~\ref{tab:cryptflow2}, Figure~\ref{fig:round_crypTFlow}, Table~\ref{tab:mpcformer}, and Figure~\ref{fig:round_crypTen}, \hawkeye can accurately profile model communication cost on different MPL frameworks. For communication size, the proportions of operators' online/offline communication size to the total online/offline communication size outputted by \hawkeye, \texttt{CrypTFlow2}, and \texttt{CrypTen} are almost the same, i.e., the proportion differences between baselines and \hawkeye are all smaller than 0.87\%.  The slight differences between the communication size profiling results outputted by \hawkeye and dynamic profiling results could be caused by the following two reasons: (1) The communication complexity of basic operations is asymptotic rather than actual. For example, \texttt{CrypTFlow2} analyzes the communication complexity upper bound of truncation and comparison operations. Because we configure the communication cost for truncation and comparison operations of \texttt{CrypTFlow2} with the upper bound rather than the actual complexity, the communication sizes outputted by \hawkeye are slightly larger than those outputted by \texttt{CrypTFlow2}  (2) The implementations of high-level operations are different. For example, \texttt{CrypTen} implements the max operation by mixing the tree-based and pairwise comparison methods, while \hawkeye purely uses the tree-based method. Thus, \texttt{CrypTen} has a larger communication size on the softmax function than \hawkeye.}

\wqruan{For communication rounds, as is shown in Figure~\ref{fig:round_crypTFlow} and Figure~\ref{fig:round_crypTen}, \hawkeye can accurately profile the communication rounds of most CNN operators on \texttt{CrypTFlow2} and all transformer operators on \texttt{CrypTen}. The main difference between the results outputted by \texttt{CrypTFlow2} and \hawkeye lies in the communication rounds of the Conv2d operators in MobileNet-V3 and ShuffleNet-V2 models. The main reason is that the implementation of the group convolution operator, which is used by MobileNet-V3 and ShuffleNet-V2, in \texttt{CrypTFlow2} is not parallel. Therefore, the group convolution operator in \texttt{CrypTFlow2} would require the group number times of communication rounds than the parallel implementation in \hawkeye. The above results show that besides helping model designers analyze model communication costs, \hawkeye could also help MPL framework developers find the performance issues in their implementation. }

\wqruan{Besides \texttt{CrypTFlow2} and \texttt{CrypTen}, we compare \hawkeye with two mixed-protocol MPL frameworks (\texttt{Delphi}~\cite{mishra2020delphi} and \texttt{Cheetah}~\cite{Cheetah}) that rely on garbled circuit (GC) and HE. Due to the space limitation, we show the experimental results in Appendix~\ref{appendix:mixed-protocol}. Furthermore, to show the accuracy of \hawkeye in secure model training scenarios, we compare \hawkeye with the two-party backend of \texttt{SecretFlow-SPU}, i.e., \texttt{SecretFlow-SEMI2K}, in Appendix~\ref{appdendix:secure_training}.}

\subsection{Efficiency of \hawkeye}
To demonstrate the efficiency of \hawkeye, \wqruan{we compare the runtimes of \hawkeye and the runtimes of \texttt{CrypTFlow2} and \texttt{CrypTen} in the experiments of Section~\ref{subsec:accuracy_garnet}.} All experiments are run on a Linux server equipped with two 32-core 2.30 GHz Intel Xeon CPUs and 512GB of RAM.

\begin{table}[ht]
\centering
\caption{\wqruan{Runtimes of static and dynamic profiling for five secure model inference processes in Section~\ref{subsec:accuracy_garnet}}.  We report the average results of five runs and show the standard deviations in brackets.}
\scalebox{0.7}{
\begin{tabular}{c|cc}
\hline
 Network                        & \multicolumn{1}{c|}{ \wqruan{Static profiling (s)}}   & \wqruan{Dynamic profiling (s)} \\ \hline
Densenet-121           & \multicolumn{1}{c|}{\wqruan{34.38 ($\pm$ 0.62)}}   & \wqruan{4,429.01 ($\pm$ 65.74)}  \\ 
Resnet-50              & \multicolumn{1}{c|}{\wqruan{39.24 ($\pm$ 0.13)}}   & \wqruan{5,698.77 ($\pm$ 80.06)}  \\ 
\wqruanother{Mobilenet-V3}           & \multicolumn{1}{c|}{\wqruan{31.47 ($\pm$ 0.31}}   & \wqruan{811.92 ($\pm$ 1.81)}  \\ 
Shufflenet-V2          & \multicolumn{1}{c|}{ \wqruan{15.50 ($\pm$ 0.12)}}   & \wqruan{319.34 ($\pm$ 3.44)}   \\ 
BERT$_{\textsc{base}}$ & \multicolumn{1}{c|}{\wqruan{471.54 ($\pm$ 4.98)}} & \wqruan{618.90 ($\pm$ 2.11)}  \\ \hline
\end{tabular}}
\label{tab:compilation_time}
\end{table}
\wqruan{As is shown in Table~\ref{tab:compilation_time}, the efficiency of \hawkeye is promising for model communication cost profiling. Concretely, \hawkeye can profile all CNN models within one minute. The profiling time is about eight minutes even for large BERT$_{\textsc{base}}$ model. In contrast, dynamic profiling the models using \texttt{CrypTFlow2} or \texttt{CrypTen} requires much more time than \hawkeye ($1.31\times \sim 145.23\times$). As a result, \hawkeye enables model designers to efficiently profile the model communication cost and design MPC-friendly models agilely.}

\subsection{Ease of Use}~\label{subsec:ease_use}
We then report on the ease of using \hawkeye to profile the communication cost of models in \texttt{PyTorch} by showing an example of modifying a \texttt{PyTorch}-based logistics regression model training codes to be \hawkeye-based.

\begin{lstlisting}[escapechar=!,mathescape,xleftmargin=2em,framexleftmargin=2em, language=python, columns=fullflexible, caption= {An example of modifying a \texttt{PyTorch}-based logistics regression model training codes to be \hawkeye-based. The removed \texttt{PyTorch} codes are highlighted on a red background and labeled with a minus sign at the beginning of the line. The newly added \hawkeye codes are highlighted on a green background and labeled with a plus sign at the beginning of the line.}, label={listing:code_example}]
class LogisticRegression(nn.Module):
    def __init__(self, n_inputs, n_outputs):
        super(LogisticRegression, self).__init__()
        self.linear = nn.Linear(n_inputs, n_outputs)
    def forward(self,x):
        out = F.sigmoid(self.linear(x))
        return out
!\colorbox[RGB]{255,235,233}{$-$ mnist = datasets.MNIST(root='./data', train=True)}!
!\colorbox[RGB]{230,255,236}{$+$ x = Tensor(60000, 784).get\_input\_from(0)}!
!\colorbox[RGB]{230,255,236}{$+$ y = Tensor(60000, 10).get\_input\_from(0)}!
!\colorbox[RGB]{255,235,233}{$-$ dataloader = DataLoader(mnist, batch\_size=128)}!
!\colorbox[RGB]{230,255,236}{$+$ dataloader = DataLoader(x, y, batch\_size = 128)}!
model = LogisticRegression(784, 10)
optimizer = optim.SGD(model.parameters(), lr = 0.01)
criterion = nn.CrossEntropyLoss()
model.train()
!\colorbox[RGB]{255,235,233}{$-$ for i in range(10):}!
!\colorbox[RGB]{230,255,236}{$+$ @for\_range(10)}!
!\colorbox[RGB]{230,255,236}{$+$ def \_(i):}!
    x, labels = dataloader[i]
    optimizer.zero_grad()
    output = model(x)
    loss = criterion(output, labels)
    loss.backward()
    optimizer.step()
\end{lstlisting}

As is shown in Listing~\ref{listing:code_example}, the model construction process, model forward process, model backward process, and model optimization process of \hawkeye-based codes are fully consistent with those of \texttt{PyTorch}-based codes. The main differences fall on the data loading and the definition of the loop: (1) Rather than loading local data, in \hawkeye, model designers need to specify the data source and then use the input data to initialize the dataloader (Line 8-12). (2) The loop interface of \hawkeye slightly differs from that of \texttt{PyTorch} (Line 17-19). The differences should be subtle. Meanwhile, because the Autograd library of \hawkeye integrates the communication cost profiling method, model designers can profile the communication cost of the secure logistics regression model training process by directly running \hawkeye to analyze the program shown in Listing~\ref{listing:code_example} without manually inserting test instruments. As a result, model designers can use \hawkeye to profile the communication cost of complex models (\wqruanother{e.g.,} transformers) in \texttt{PyTorch} by modifying less than ten lines of code. In contrast, Li et al.~\cite{li2023mpcformer} have to manually insert about 100 lines of codes to dynamically profile the communication cost of secure transformer inference processes on \texttt{CrypTen}. More examples of implementing complex models in \hawkeye can be found in our source codes.

\subsection{Case Studies}~\label{subsec:case_studies}
In this section, we conduct three case studies to show the practical applications of \hawkeye. Firstly, to show that \hawkeye can help model designers find communication bottlenecks of models on MPL frameworks with different security models, we apply \hawkeye to profile model communication cost on four MPL frameworks whose security models are different. Secondly, to show that \hawkeye can help model designers choose a proper optimizer for secure model training, we apply \hawkeye to profile the communication cost of secure model training processes with two different optimizers. Finally, we apply \hawkeye in model computational graph optimization to improve the efficiency of secure model inference.

\subsubsection{Communication bottlenecks of models on MPL frameworks with different security models}
\noindent\textbf{Security models of MPL frameworks.} We first introduce security models of MPL frameworks. The security model of an MPL framework refers to assumptions the MPL framework makes about parties. In scenarios with different security requirements, model designers usually need to employ MPL frameworks with the corresponding security models. One security model generally compromises two dimensions: the behavior of the parties and the number of colluded parties. Depending on whether parties follow the protocols, the security models of MPL frameworks can be classified as semi-honest and malicious. Depending on whether the number of colluded parties is strictly below half of the total number of parties or not, the security models of MPL frameworks can be classified as honest-majority and dishonest-majority.

We apply \hawkeye to profile the secure inference processes of Resnet-50 and BERT$_{\textsc{base}}$ models on four MPL frameworks, \wqruanother{i.e.,}  \texttt{ABY}~\cite{aby}, \texttt{SPDZ-2k}~\cite{spdz2k}, \texttt{ABY3}~\cite{mohassel2018aby3}, and \texttt{Falcon}~\cite{wagh2020falcon}, whose security models are (semi-honest, dishonest-majority), (malicious, dishonest-majority), (semi-honest, honest-majority),  (malicious, honest-majority) respectively.
For parameter settings, we set the bit length as $64$, the statistical security parameter as $40$, the computational security parameter as $128$, the bit length of fixed-point numbers' fractional part as $16$, and the number of parties as two for \texttt{ABY} and \texttt{SPDZ-2k}, three for \texttt{ABY3} and \texttt{Falcon}. The input data sizes are consistent with those used in Section~\ref{subsec:accuracy_garnet}.
% We show communication cost configurations of the four MPL frameworks in Appendix~\ref{appendix:protocol_config}.

\begin{figure}[ht]
    \centering
    \includegraphics[width=0.4\textwidth]{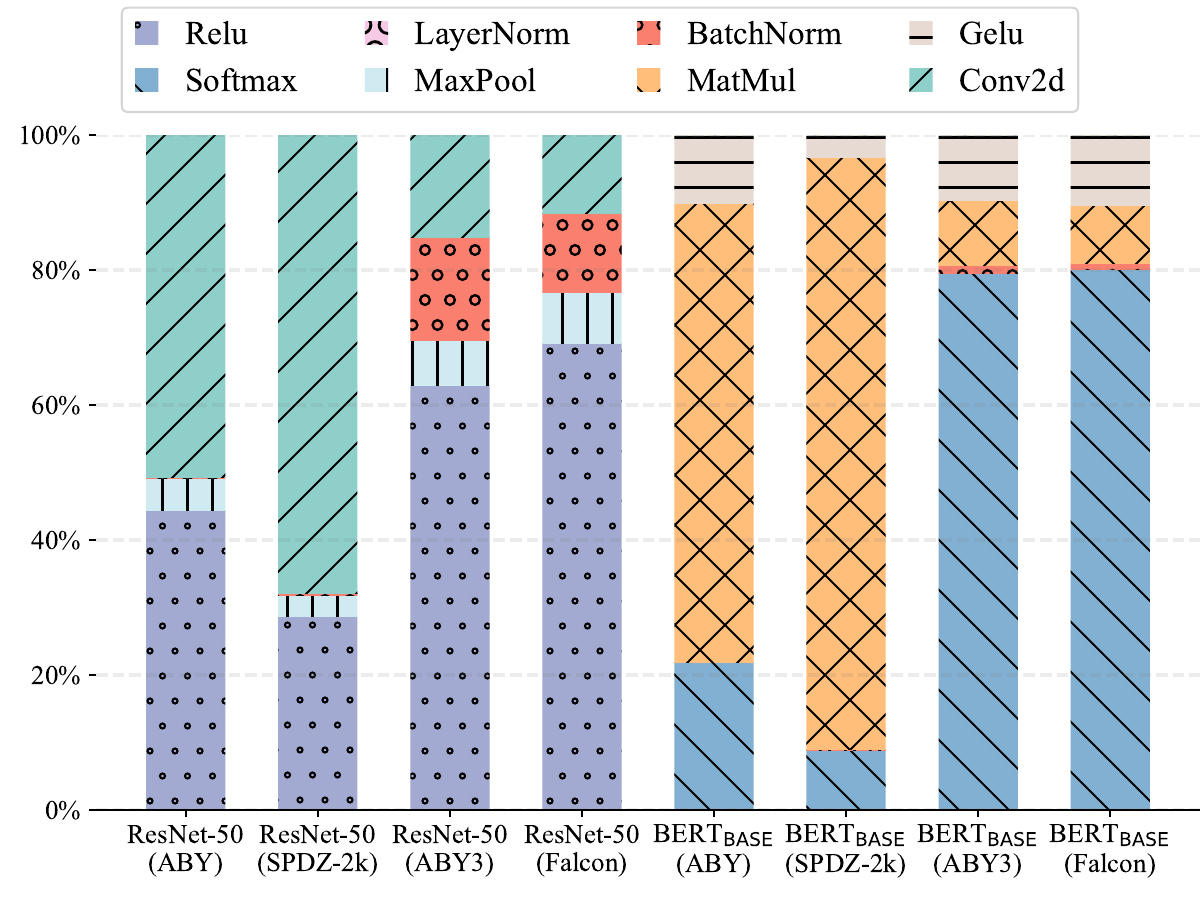}
    \caption{The proportion of each operator's online communication size to the total online communication size on four MPL frameworks with different security models.}
    \label{fig:protocol_profiling}
\end{figure}

As is shown in Figure~\ref{fig:protocol_profiling}, two dimensions of the security model have significantly different impacts on communication bottlenecks of models: (1) Under the same assumption on the number of colluded parties, switching the assumption on the behavior of parties from semi-honest to malicious slightly changes the profiling results. Therefore, MPC-friendly models optimized for semi-honest MPL frameworks could remain effective for malicious MPL frameworks.  (2) When underlying MPL frameworks are designed for a dishonest majority, the communication bottlenecks are linear operators, \wqruanother{i.e.,} Matmul or Conv2d. In contrast, the communication bottlenecks become non-linear operators (\wqruanother{i.e.,} Softmax, Relu, Gelu, MaxPool) when underlying MPL frameworks are designed for an honest majority. Therefore, model designers would need to tailor MPC-friendly models for different scenarios where the assumptions on the number of colluded parties are different. 
The above results show that \hawkeye can help model designers efficiently find communication bottlenecks of models on MPL frameworks with different security models. 

\subsubsection{Choice of optimizers in secure model training}  We apply \hawkeye to profile the communication cost of secure model training processes with two optimizers. A few studies~\cite{10.1145/3411501.3419427,DBLP:journals/popets/AttrapadungHIKM22} show that in secure model training, Adam~\cite{adam} could be a better optimizer than SGD because Adam can significantly improve the convergence speed. However, Adam incurs much more communication overhead than SGD. To quantitatively analyze the communication cost of optimizers, we apply \hawkeye to profile the communication cost of three secure CNN model training processes (LeNet, AlexNet, VGG-16) on two MPL frameworks (\texttt{ABY}, \texttt{ABY3}). Meanwhile, following previous studies~\cite{watson22piranha,cryptGPU}, we replace MaxPool in the CNN models with AvgPool. Finally, we set the input data size as $128 \times 3 \times 28 \times 28$ for LeNet, $128 \times 3 \times 224 \times 224$ for AlexNet and VGG-16, where $128$ is the batch size. Other parameter settings are the same as those of the first case study. Note that we run the secure model training processes on one batch of data.  Because the model training process is the same for each batch of data, the proportion of each part's online communication size to the total online communication size remains unchanged as the batch number changes.

\begin{table}[ht]
\centering
\caption{The online communication size of gradient computation and optimization of three secure CNN model training processes. We report the proportion of each part's online communication size to the total online communication size and the online communication size of each part.}
\scalebox{0.7}{
\begin{tabular}{c|c|cc}
\hline
Framework              & Model        & Grad Computation & Optimization \\ \hline
\multirow{6}{*}{\texttt{ABY}~\cite{aby}}  & Lenet-SGD    & 100.00\% (10.47GB)               & 0.00\% (0GB)                       \\ 
                      & Lenet-Adam   & 92.17\% (10.47GB)                      & 7.83\% (0.89GB)                  \\ 
                      & AlexNet-SGD  & 100.00\% (12,526.23GB)                   & 0.00\% (0GB)                       \\ 
                      & AlexNet-Adam & 97.06\% (12,526.23GB)                   & 2.94\% (379.67GB)                \\ 
                      & VGG-16-SGD   &  100.00\% (189,079.51GB)                  & 0.00\% (0GB)                       \\  
                      & VGG-16-Adam  & 99.55\% (189,079.51GB)                  & 0.45\% (859.72GB)                \\ \hline
\multirow{6}{*}{\texttt{ABY3}~\cite{mohassel2018aby3}} & Lenet-SGD    & 98.94\% (0.10GB)                   & 1.06\% ($<0.01$GB)         \\ 
                      & Lenet-Adam   & 32.26\% (0.10GB)                      & 67.74\% (0.21GB)                  \\ 
                      & AlexNet-SGD  & 96.34\% (12.12GB)                     & 3.66\% (0.46GB)                  \\ 
                      & AlexNet-Adam & 3.99\% (12.12GB)                     & 96.01\% (291.35GB)                \\ 
                      & VGG-16-SGD   & 99.80\% (521.60GB)                    & 0.20\% (1.03GB)                 \\ 
                      & VGG-16-Adam  & 44.15\% (521.60GB)                    & 55.85\% (659.74GB)                \\ \hline
\end{tabular}
}
\label{tab:choice_optimizer}
\end{table}

As is shown in Table~\ref{tab:choice_optimizer}, the extra communication cost incurred by Adam significantly differs among different models and MPL frameworks. When training models on \texttt{ABY}, the online communication size of Adam only accounts for $0.45\% \sim 7.83\%$ of the total online communication size. In this case, replacing SGD with Adam would significantly improve the training efficiency because the extra communication cost incurred by Adam is minor compared with the communication cost of gradient computation. In contrast, when training models on \texttt{ABY3}, replacing SGD with Adam would increase the total online communication size by $2.26 \sim 24.12$ times. Especially when securely training the AlexNet model on \texttt{ABY3}, replacing SGD with Adam would cause the total online communication size to increase by $24.12$ times. In this case, SGD should be a better optimizer than Adam because the convergence speed improvement brought by Adam cannot cover its extra communication cost. 

\subsubsection{\wqruan{Computational graph optimization}}
\wqruan{To further show the practical application of \hawkeye, we combine \hawkeye with TASO~\cite{10.1145/3341301.3359630}, a classical computational graph optimization method for deep learning models, to effectively reduce the communication overhead of secure model inference. TASO improves the efficiency of secure model inference by changing the structure of the computational graphs that represent the model inference process. Meanwhile, TASO ensures that the optimized computational graph is equivalent to the original computational graph. In this experiment, we use the online communication size outputted by \hawkeye as the cost model of TASO. For the underlying MPL framework, we use the three-party protocol of \texttt{Deep-MPC} proposed by Keller and Sun~\cite{pmlr-v162-keller22a}, whose source codes are included in \texttt{MP-SPDZ}, as our target MPL framework. We show the communication cost configuration of \texttt{Deep-MPC} in Appendix~\ref{appendix:protocol_config}. Meanwhile, following Jia et al.~\cite{10.1145/3341301.3359630}, we choose the backbone network (i.e., model components excluding the stem component and the final classifier) of ResNet-18 and ResNet-50 models as target models.}

\begin{table}[ht]
\centering
\caption{\wqruan{The communication size and communication time of two CNN models that are optimized by original TASO and \hawkeye-enhanced TASO (TASO-\hawkeye). The communication time is obtained under the WAN setting where round-trip time is 72ms, and bandwidth is 9 MBps. We report the average results of five runs and show the standard deviations in brackets.}}
\scalebox{0.7}{
\begin{tabular}{c|c|c|c}
\hline
Model                      & Method       & Comm Size (MB) & Comm Time (s)    \\ \hline
\multirow{3}{*}{ResNet-18} & Original     & \wqruan{286.10MB}       & \wqruan{41.30s ($pm$ 0.68s)}   \\ \cline{2-4} 
                           & TASO   &  \wqruan{247.49MB }      &  \wqruan{37.00s ($pm$ 0.74s)}   \\ \cline{2-4} 
                           & TASO-\hawkeye &  \wqruan{210.92MB}       &  \wqruan{30.54s ($pm$ 1.26s) }  \\ \hline
\multirow{3}{*}{ResNet-50} & Original     &  \wqruan{ 1758.85MB }     &  \wqruan{207.71s ($pm$ 8.17s)} \\ \cline{2-4} 
                           & TASO   &  \wqruan{1647.08MB }     &  \wqruan{193.54s ($pm$ 8.72s)}  \\ \cline{2-4} 
                           & TASO-\hawkeye &  \wqruan{1492.64MB  }    &  \wqruan{174.27s ($pm$ 5.48s)}  \\ \hline
\end{tabular}}
\label{tab:taso}
\end{table}
\wqruan{As is shown in Table~\ref{tab:taso}, \hawkeye-enhanced TASO significantly outperforms the original TASO. Concretely, \hawkeye-enhanced TASO can reduce the communication overhead by 15.14\% $\sim$ 26.28\% and the communication time by 16.10\% $\sim$ 26.05\%, which is 1.95 $\sim$ 2.38 times and 2.36 $\sim$ 2.50 times that of the original TASO, respectively. The above results show that \hawkeye can effectively help the efficiency optimization of secure model inference.}

%% file: tex/6.RelatedWork.tex
\section{Related Work}~\label{sec:related}
\noindent\textbf{Design and optimization of MPC-friendly models.} Many studies~\cite{ganesan2022efficient, Peng_2023_ICCV, li2023mpcformer, liu2023llms, MPCViT, dhyani2023privit,10179422} try designing and optimizing MPC-friendly models by dynamically profiling model communication cost on one or two specific MPL frameworks~\cite{9444564}. Ganesan et al.~\cite{ganesan2022efficient} design an MPC-friendly convolution operator to optimize the efficiency of secure CNN model inference on \texttt{CrypTFlow2}~\cite{rathee2020cryptflow2}. Subsequently, Peng et al.~\cite{Peng_2023_ICCV} propose AutoReP to automatically replace part of Relu functions of CNN models with polynomial functions to speed up secure CNN model inference on \texttt{CrypTen}~\cite{crypten2020}. In addition to CNN models, Liu et al.~\cite{liu2023llms} design an MPC-friendly transformer model by profiling the communication cost of secure transformer inference on \texttt{Cheetah}~\cite{Cheetah} and \texttt{CrypTFlow2}~\cite{rathee2020cryptflow2}. At the same time, Li et al.~\cite{li2023mpcformer} design an MPC-friendly variant of the BERT model, namely MPCFormer, by profiling the communication cost of secure BERT$_{\textsc{base}}$ model inference process on \texttt{CrypTen}~\cite{crypten2020}. For the image classification task based on the transformer model, Zeng et al.~\cite{MPCViT} propose an MPC-friendly vision transformer (ViT)~\cite{dosovitskiy2021an} model, namely MPCViT, by profiling the communication cost of secure ViT inference process on \texttt{SecretFlow-SPU}~\cite{secretflow}. In addition, Dhyani et al.~\cite{dhyani2023privit} design another MPC-friendly ViT model by profiling the communication cost of the secure ViT inference process on \texttt{DELPHI}~\cite{244032}. The above studies optimize the structure of MPC-friendly models by dynamically profiling model communication cost on one or two specific MPL frameworks. Therefore, their optimizations may only be effective in limited scenarios. With \hawkeye, model designers can statically profile model communication cost on multiple MPL frameworks without implementing specific models on these MPL frameworks without dynamic profiling, thus significantly improving the automation of the optimization of MPC-friendly models.

% \smallskip
\noindent\textbf{Cost models for hybrid protocol compilers.}  Several studies~\cite{costco, HyCC, OPA, CheapSMC,10179397} try modeling the cost of different MPC protocols to select an MPC protocol combination for efficient MPC. \texttt{CheapSMC}~\cite{CheapSMC} models the cost of an MPC protocol by running every basic operation of the MPC protocol to obtain the cost of each basic operation. Then, they estimate the cost of an input program by counting the number of operations required by the input program. \texttt{OPA}~\cite{OPA} and \texttt{HyCC}~\cite{HyCC} further improve \texttt{CheapSMC} by considering the circuit structure in their cost model. Concretely, in addition to testing the runtime of every single operation, they test circuits with different levels of parallelism, \wqruanother{i.e.,} computing multiple operations in one communication round. Beyond its previous studies, \texttt{CostCO}~\cite{costco} automatically generates and tests thousands of circuits with different structures to estimate the cost of MPC protocols. These studies aim to enhance the existing hybrid protocol compilers to find an efficient combination of MPC protocols for a given secure computation task. Unlike the above studies, \hawkeye aims to help model designers optimize the structures of MPC-friendly models for MPL frameworks, thus improving the efficiency of MPL from the machine learning perspective. Therefore, \hawkeye and existing studies on cost models for hybrid protocol compilers are orthogonal.

%% file: tex/7.Discussion.tex
\section{Discussion}~\label{sec:dis}
\noindent\textbf{Burdensome human efforts reduction by \hawkeye.} \hawkeye can reduce the burdensome human efforts of dynamic profiling from two aspects: (1) model designers can statically profile the communication cost of one model on multiple MPL frameworks without manually implementing the model in the MPL frameworks. Thus, the static profiling process in \hawkeye can significantly reduce human efforts. (2) \hawkeye offers an Autograd library whose model construction interfaces are fully consistent with those of \texttt{PyTorch} and integrates our proposed static communication cost profiling method. The Autograd library of \hawkeye can significantly reduce human efforts of implementing models in \hawkeye and inserting test instruments.

% \smallskip
\noindent\textbf{Local computation cost optimization.}
Local computation cost optimization~\cite{watson22piranha, cryptGPU} is another important research topic in the field of MPL. However, because communication cost is still the main performance bottleneck of MPL, we mainly consider profiling the communication cost of models in this paper. Watson et al.~\cite{watson22piranha} also show that with the acceleration of GPU, even in the LAN setting, the local computation time is significantly smaller than the communication time. In the WAN setting, the local computation time is negligible compared with the communication time. Therefore, the communication cost is the main optimization goal of studies on the optimization of MPC-friendly models~\cite{li2023mpcformer, ganesan2022efficient, MPCViT}. In addition, the communication optimization and local computation optimization are orthogonal. These two optimizations can improve the efficiency of MPL in different ways.

% \smallskip
\noindent\textbf{\wqruan{The adaptive protocol assignment for mixed-protocol frameworks.}} \wqruan{The mixed-protocol MPL frameworks (i.e., \texttt{CrypTen}, \texttt{Cheetah}, \texttt{Deep-MPC} and \texttt{Delphi}) involved in our experiments use static protocol assignments introduced in Section~\ref{subsec:label_interface}, i.e., model designers specify the protocol assignment and configure the communication cost of corresponding non-linear operations in \texttt{HawkEye}.}

\wqruan{However, emerging mixed-protocol MPL frameworks (e.g., \texttt{Silph}~\cite{10179397}) can adaptively assign protocols to efficiently perform non-linear operations according to circuit structures. \hawkeye can be extended to profile the communication cost of these frameworks that employ the adaptive protocol assignment when model designers provide protocol assignment strategies. Concretely, we can deliver the circuit structure information (e.g., the number of non-linear operators to be computed in parallel) in the parameters of the operation communication cost functions. Then, model designers can adaptively adjust the communication cost of non-linear operations according to the delivered circuit structure information.}
% Therefore, model designers can use \hawkeye to accurately profile the communication costs of mixed-protocol frameworks that employ the adaptive protocol assignment.   }

%% file: tex/8.Conclusion.tex
\section{Conclusion and Future Work}~\label{sec:conclu}
In this paper, we propose \hawkeye, a static communication cost profiling framework to enable model designers to \wqruanother{get} the accurate communication cost of models in MPL frameworks without dynamic profiling. \hawkeye contributes two folders: a static communication cost profiling method and an Autograd library. The static communication cost profiling method statically profiles model communication cost by breaking high-level components as compositions of basic operations. Meanwhile, the Autograd library has model construction interfaces that are fully consistent with those of \texttt{PyTorch} and integrates the static communication cost profiling method to profile the communication cost of models in \texttt{PyTorch}. Finally, we conduct a series of experiments to compare the static profiling results outputted by \hawkeye with the dynamic profiling results from two MPL frameworks. The experimental results show that \hawkeye can accurately profile the model communication cost. As a result, \hawkeye can effectively help model designers optimize the structures of MPC-friendly models in MPL frameworks.

In the future,  we will extend \hawkeye to \wqruanother{profile the communication costs of MPL frameworks that employ adaptive protocol assignments} and the local computation cost of models in MPL frameworks, such that model designers can obtain more comprehensive information when they design and optimize MPC-friendly models.

%% file: tex/9.appendix.tex
\section{Communication Cost Configurations of MPL Frameworks}~\label{appendix:protocol_config}
\wqruan{In this section, we introduce how we configure the communication costs for basic operations of ten MPL frameworks (i.e., \texttt{CrypTFlow2}~\cite{rathee2020cryptflow2}, \texttt{CrypTen}~\cite{crypten2020}, \texttt{ABY}~\cite{aby}, \texttt{SPDZ-2k}~\cite{spdz2k}, \texttt{ABY3}~\cite{mohassel2018aby3}, and \texttt{Falcon}~\cite{wagh2020falcon}, \texttt{Delphi}~\cite{mishra2020delphi}, \texttt{Cheetah}~\cite{Cheetah}, \texttt{Deep-MPC}~\cite{pmlr-v162-keller22a}, \texttt{SecretFLow-SEMI2K}~\cite{secretflow}) involved in our experiments. We first introduce the extra parameter for some basic operations. Secure multiplication (`muls') has three extra parameters: the degree of HE polynomial $\mathit{deg}$, HE prime coefficients modulus $\mathit{mod}$, and $\mathit{size}$ that indicates the number of secure multiplications executed in parallel. Matrix multiplication (`matmuls') has five extra parameters: the row number of the first input matrix $p$, the column number of the first input matrix $q$, the column number of the second input matrix $r$, the degree of HE polynomial $\mathit{deg}$, HE prime coefficients modulus $\mathit{mod}$. Secure truncation (`TruncPr') has one extra parameter: $\mathit{knownmsb}$ that indicates whether the most significant bit (msb) of the input is known.} 

\texttt{CrypTFlow2}~\cite{rathee2020cryptflow2}: We configure the communication costs for basic operations of \texttt{CrypTFlow2}~\cite{rathee2020cryptflow2} by analyzing Tables 1 and 2, Section 4.2, and Appendix F of \cite{rathee2020cryptflow2} and running its source codes. \wqruan{\texttt{CrypTFlow2} integrates an optimization for truncation proposed by Rathee et al.~\cite{sirnn}. Therefore, we configure two versions of the communication cost of `TruncPr' for \texttt{CrypTFlow2}. Besides, for the communication round of secure matrix multiplication, we find that \texttt{CrypTFlow2} would partition the transferred messages for large input messages. Therefore, we set the communication round of `matmuls' according to the partition strategy defined in lines 239-248 of `linear-ot.cpp' in the source codes of \texttt{CrypTFlow2}.} Because \texttt{CrypTFlow2} finishes all computations in the online phase, we set the communication cost of its offline phases as 0. The communication cost configuration of \texttt{CrypTFlow2} is shown in Listing~\ref{list:cryptflow2_cost}. Note that we use ring-based \texttt{CrypTFlow2} following Ganesan et al.~\cite{ganesan2022efficient}.

\begin{lstlisting}[mathescape,xleftmargin=2em,framexleftmargin=2em, caption = {Communication cost configuration for basic operations of $\texttt{CrypTFlow2}$~\cite{rathee2020cryptflow2}.}, columns=fullflexible, label = {list:cryptflow2_cost}]
class CrypTFlow2(Cost):
  cost_func_dict = {
    "share": lambda $k$, $\kappa_s$, $\kappa$, $f$, $m$: (0, 0, 0, 0),
    "reveal": lambda $k$, $\kappa_s$,$\kappa$, $f$, $m$: (2*$k$, 1, 0, 0),
    "muls": lambda $k$, $\kappa_s$,$\kappa$, $f$, $m$, $\mathit{size}$, $\mathit{deg}$, $\mathit{mod}$: 
    ($\mathit{size}$*$k$*($\lceil$($k$+1)/2$\rceil$+$\kappa$), 
    2, 0, 0),
    "matmuls": lambda $k$, $\kappa_s$, $\kappa$, $f$, $m$, $p$, $q$, $r$, $\mathit{deg}$, $\mathit{mod}$: 
    ($q$*$r$*$k$*($p$*$\lceil$($k$+1)/2$\rceil$+$\kappa$), 2*$k$/ ($\lceil$ $2^{24}$/($p$*$q$*$r$)$\rceil$),
    0, 0) ,
    "TruncPr": lambda $k$, $\kappa_s$, $\kappa$, $f$, $m$, $\mathit{knownmsb}$: 
    (($\kappa$+14)*$f$+ 2*$\kappa$ + 4*$k$ if $\mathit{knownmsb}$ else $\kappa$*($k$+2)+19*$k$
    +($\kappa$+14)*$f$, 2 if $\mathit{knownmsb}$ else 2*log($k$)+2, 0, 0) ,    
    "LTZ": lambda $k$, $\kappa_s$, $\kappa$, $f$, $m$:  
    (($\kappa$+18)*$k$, log($k$), 0, 0) 
    }
\end{lstlisting}

\begin{lstlisting}[mathescape,xleftmargin=2em,framexleftmargin=2em, caption = {Communication cost configuration for basic operations of $\texttt{CrypTen}$~\cite{crypten2020}.}, columns=fullflexible, label = {list:crypten_cost}]
class CrypTen(Cost):
  cost_func_dict = {
    "share": lambda $k$, $\kappa_s$, $\kappa$, $f$, $m$: (0, 0, 0, 0),
    "reveal": lambda $k$, $\kappa_s$,$\kappa$, $f$, $m$: (2*$k$, 1, 0, 0),
    "muls": lambda $k$, $\kappa_s$,$\kappa$, $f$, $m$, $\mathit{size}$, $\mathit{deg}$, $\mathit{mod}$: 
    (2*$k$*$\mathit{size}$, 1, $k$*$\mathit{size}$, 3),
    "matmuls": lambda $k$, $\kappa_s$, $\kappa$, $f$, $m$, $p$ ,$q$, $r$, $\mathit{deg}$, $\mathit{mod}$: 
    (($p$*$q$+$q$*$r$)*$k$*2, 1, $p$*$r$*$k$, 3) ,
    "TruncPr": lambda $k$, $\kappa_s$, $\kappa$, $f$, $m$, $\mathit{knownmsb}$: (0, 0, 0, 0),    
    "LTZ": lambda $k$, $\kappa_s$, $\kappa$, $f$, $m$:  
    (54*$k$, log($k$)+2, 14*$k$, (log($k$)+2)*3), 
    "exp_fx": lambda $k$, $\kappa_s$, $\kappa$, $f$, $m$:  (16*$k$, 8, 8*$k$, 24), 
    "EQZ": lambda $k$, $\kappa_s$, $\kappa$, $f$, $m$:  (26*$k$, log($k$), 7*$k$, 21), 
    "Reciprocal" : lambda $k$, $\kappa_s$, $\kappa$, $f$, $m$:  
    (138*$k$, 38, 44*$k$, 114)
    }
\end{lstlisting}

\texttt{CrypTen}~\cite{crypten2020}:\wqruan{We configure the communication costs for basic operations of the two-party backend of \texttt{CrypTen}~\cite{crypten2020} by analyzing the protocol description in Appendices C.1, C.2 of \cite{crypten2020}.}  Furthermore, we update the communication cost configuration by running these basic operations with open-source codes of \texttt{CrypTen}. The communication cost configuration of \texttt{CrypTen} is shown in Listing~\ref{list:crypten_cost}. \wqruan{For its offline communication costs, we configure them by running each basic operation under the setting of one trusted dealer, i.e., trusted third party (TTP).}

\begin{lstlisting}[mathescape,xleftmargin=2em,framexleftmargin=2em, caption = {Communication cost configuration for basic operations of \texttt{ABY}~\cite{aby}.}, columns=fullflexible, label = {list:aby_cost}]
class ABY(Cost):
  cost_func_dict = {
    "share": lambda $k$, $\kappa_s$, $\kappa$, $f$, $m$: (0, 0, 0, 0),
    "reveal": lambda $k$, $\kappa_s$,$\kappa$, $f$, $m$: (2*$k$, 1, 0, 0),
    "muls": lambda $k$, $\kappa_s$,$\kappa$, $f$, $m$, $\mathit{size}$, $\mathit{deg}$, $\mathit{mod}$: 
    (4*$k$*$\mathit{size}$, 1, 
    (2*$\kappa$+$k$+1)*$k$*$\mathit{size}$, 2),
    "matmuls": lambda $k$, $\kappa_s$, $\kappa$, $f$, $m$, $p$ ,$q$, $r$, $\mathit{deg}$, $\mathit{mod}$: 
    ($p$*$q$*$r$*$k$*4, 1, $p$*$q$*$r$*(2*$\kappa$+$k$+1)*$k$, 2),
    "TruncPr": lambda $k$, $\kappa_s$, $\kappa$, $f$, $m$, $p$ ,$q$, $r$, $\mathit{knownmsb}$: 
    (0, 0, 0, 0),    
    "LTZ": lambda $k$, $\kappa_s$, $\kappa$, $f$, $m$, $p$ ,$q$, $r$:  
    ($\kappa$*$k$*7+($$k$^2$+$k$)/2, 4, 5*$\kappa$*$k$, 2) 
    }
\end{lstlisting}

\texttt{ABY}~\cite{aby}: \wqruan{We configure the communication costs for basic operations of \texttt{ABY} according to Section III-A, Table1 of \cite{aby}, and Algorithm 4 of \cite{pmlr-v162-keller22a}. We configure the communication cost for `reveal' and `muls' of \texttt{ABY}~\cite{aby} according to the protocol description and complexity analysis of Section III-A of \cite{aby}. For the communication cost of `matmuls', we directly use $p$*$q$*$r$ multiplications to implement it because \texttt{ABY} does not have special optimization for matrix multiplication. For the communication cost of `TruncPr', we use the local truncation of \texttt{SecureML}. For `LTZ', we implement it in \texttt{ABY} by calling one A2Y, one Y2B, and one B2A in \texttt{ABY}. Then we configure the communication cost of `LTZ' according to Table 1 of \cite{aby}.  Besides, the communication cost of `share' is from the optimized protocol in Algorithm 4 of \cite{pmlr-v162-keller22a}.} The communication cost configuration of \texttt{ABY} is shown in Listing~\ref{list:aby_cost}.

\texttt{SPDZ-2k}~\cite{spdz2k}: We configure the communication costs for basic operations of \texttt{SPDZ-2k}~\cite{spdz2k} according to Figures 6, 9, 11, 12 and 13, and Section 7 of \cite{spdz2k}. 

\wqruan{Firstly, we analyze the communication costs of two basic protocols: authentication and triple generation. According to Figure 11 and the complexity analysis in Section 7 of \cite{spdz2k}, the authentication protocol requires $\mathit{\kappa_s*max(k+\kappa_s, 2*\kappa_s)*m*(m-1)}$ bits and 3 rounds (2 for authentication and 1 for consistency check). For simplicity, we assume that $k$ is always larger than $\kappa_s$. According to Figures 12, 13 and the complexity analysis in Section 7 of \cite{spdz2k}, the triple generation protocol requires $\mathit{18*\kappa_s^2+4*k^2+17*\kappa*k)*m*(m-1)}$ bits and 8 rounds (2 for multiplication and 6 for authentication and sacrifice). we omit the communication costs for batch check of triples that are negligible compared to those of triple generation. }

\begin{lstlisting}[mathescape,xleftmargin=2em,framexleftmargin=2em, caption = {Communication cost configuration for basic operations of \texttt{SPDZ-2k}~\cite{spdz2k}.}, columns=fullflexible, label = {list:spdz_cost}]
class SPDZ-2k(Cost):
  cost_func_dict = {
    "share": lambda $k$, $\kappa_s$, $\kappa$, $f$, $m$: 
    (($\kappa_s$+$k$)*($m$-1), 1, ($\kappa_s$*($k$+$\kappa_s$)*$m$*($m$-1), 3),
    "reveal": lambda $k$, $\kappa_s$,$\kappa$, $f$, $m$:
    (($\kappa_s$+$k$)*$m$*($m$-1), 1,  ($\kappa_s$*($k$+$\kappa_s$)*$m$*($m$-1), 3),
    "muls": lambda $k$, $\kappa_s$,$\kappa$, $f$, $m$, $\mathit{deg}$, $\mathit{mod}$: 
    (($k$ + $\kappa_s$)*$m$*($m$-1)*2, 1, 
    (18*$\kappa_s^2$+4*$$k$^2$+17*$\kappa_s$*$k$)*$m$*($m$-1), 8),
    "matmuls": lambda $k$, $\kappa_s$, $\kappa$, $f$, $m$, $p$ ,$q$, $r$, $\mathit{deg}$, $\mathit{mod}$: 
    (($k$ + $\kappa_s$)*$m$*($m$-1)*2*$p$*$q$*$r$, 1,
    ((18*$\kappa_s^2$+4*$$k$^2$+17*$\kappa_s$*$k$))*$m$*($m$-1)*$p$*$q$*$r$, 8),
    "TruncPr": lambda $k$, $\kappa_s$, $\kappa$, $f$, $m$:
    (($k$ + $\kappa_s$)*$m$*($m$-1), 1, 
    $k$*(($\kappa_s$+$k$)*(3m+1)*($m$-1)+$\kappa_s$*($k$+$\kappa_s$)*$m$*($m$-1)*2
    +(18*$\kappa_s^2$+4*$$k$^2$+17*$\kappa_s$*$k$)*$m$*($m$-1)), 11) 
    }
\end{lstlisting}

\wqruan{Secondly, we configure the communication costs for basic operations as follows. For the communication costs of `share' and `reveal', according to Figures 6 and 9 of \cite{spdz2k}, their online phases need to broadcast one message, and the offline phases require one call of the protocol for authentication. Therefore, their online communication cost is $\mathit{(\kappa_s+k)*(m-1)}$ and $\mathit{(\kappa_s+k)*m*(m-1)}$ respectively, and offline communication cost is the communication cost of the authentication protocol. For the communication cost of `muls', according to Figures 6 and 9 of \cite{spdz2k}, its online phase requires two calls of reveal protocol to reveal masked input, and the offline phase generates a beaver triple.  Therefore, its online communication cost is $\mathit{(\kappa_s+k)*m*(m-1)*2}$, and offline communication cost is the communication cost of the triple generation protocol. The communication cost of `matmuls' is obtained by simply multiplying the communication cost of `muls' by $p*q*r$. Finally, for the communication cost of TruncPr operation, we configure its communication cost according to the description of Appendix IX.C in the study~\cite{8835310}. Its online phase requires one call of the reveal protocol, and the offline phase requires $k$ calls of the random bit generation protocol that includes one share, one reveal, and one multiplication. The communication cost configuration of \texttt{SPDZ-2k} is shown in Listing~\ref{list:spdz_cost}.}

\texttt{ABY3}~\cite{mohassel2018aby3}: We configure the communication costs for basic operations of \texttt{ABY3}~\cite{mohassel2018aby3} according to Tables 2, Figure 2, and Sections 3.2.1 and 5.2 of \cite{mohassel2018aby3}. \wqruan{For the communication costs of `share', `reveal' and `muls' operations, we obtain their communication costs by analyzing the protocol description in Section 3.2.1 of \cite{mohassel2018aby3}. For the communication cost of `matmuls', we configure it by analyzing the protocol description in Section 5.2 of \cite{mohassel2018aby3}.  We configure the communication cost of `TruncPr' by analyzing the protocol description in Figure 2 of \cite{mohassel2018aby3}. Finally, for the communication cost of `LTZ', \texttt{ABY3} implements it by calling a bit extraction and a B2A. We then set the communication cost of `LTZ' according to Table 2 of \cite{mohassel2018aby3}.} The communication cost configuration of \texttt{ABY3} is shown in Listing~\ref{list:aby3_cost}. 

\begin{lstlisting}[mathescape,xleftmargin=2em,framexleftmargin=2em, caption = {Communication cost configuration for basic operations of \texttt{ABY3}~\cite{mohassel2018aby3}.}, columns=fullflexible, label = {list:aby3_cost}]
class ABY3(Cost):
  cost_func_dict = {
    "share": lambda $k$, $\kappa_s$, $\kappa$, $f$, $m$: (3*$k$, 1, 0, 0),
    "reveal": lambda $k$, $\kappa_s$,$\kappa$, $f$, $m$: (3*$k$, 1, 0, 0),
    "muls": lambda $k$, $\kappa_s$,$\kappa$, $f$, $m$, $\mathit{size}$, $\mathit{deg}$, $\mathit{mod}$: 
    (3*$k$*$\mathit{size}$, 1, 0, 0),
    "matmuls": lambda $k$, $\kappa_s$, $\kappa$, $f$, $m$, $p$ ,$q$, $r$, $\mathit{deg}$, $\mathit{mod}$: 
    (3*$p$*$r$*$k$, 1, 0, 0),
    "TruncPr": lambda $k$, $\kappa_s$, $\kappa$, $f$, $m$, $\mathit{knownmsb}$:
    ($k$, 1, 0, 0),    
    "LTZ": lambda $k$, $\kappa_s$, $\kappa$, $f$, $m$:  (9*$k$, log($k$)+2, 0, 0) 
    }
\end{lstlisting}

\begin{lstlisting}[mathescape,xleftmargin=2em,framexleftmargin=2em, caption = {Communication cost configuration for basic operations of \texttt{Falcon}~\cite{wagh2020falcon}.}, columns=fullflexible, label = {list:falcon_cost}]
class Falcon(Cost):
  cost_func_dict = {
    "share": lambda $k$, $\kappa_s$, $\kappa$, $f$, $m$: (3*$k$, 1, 0, 0),
    "reveal": lambda $k$, $\kappa_s$,$\kappa$, $f$, $m$: (6*$k$, 1, 0, 0),
    "muls": lambda $k$, $\kappa_s$,$\kappa$, $f$, $m$, $\mathit{size}$, $\mathit{deg}$, $\mathit{mod}$: 
    (6*$k$*$\mathit{size}$, 1, 0, 0),
    "matmuls": lambda $k$, $\kappa_s$, $\kappa$, $f$, $m$, $p$ ,$q$, r, $\mathit{deg}$, $\mathit{mod}$: 
    (6*$p$*r*$k$, 1, 0, 0),
    "TruncPr": lambda $k$, $\kappa_s$, $\kappa$, $f$, $m$, $\mathit{knownmsb}$: 
    (2*$k$, 1, (6+log($k$))*$k$+(6+$\lceil$log($k$-$f$)$\rceil$)*($k$-$f$), log($k$)+2),    
    "LTZ": lambda $k$, $\kappa_s$, $\kappa$, $f$, $m$:  
    (24*$k$, log($k$)+5,(k+8+log($k$))*$k$*3, 4+2*log($k$)),
    "Pow2":  lambda $k$, $\kappa_s$, $\kappa$, $f$, $m$:  
    (24*$k$*$k$, (log($k$)+5)*$k$, 
    ((k+8+log($k$))*$k$*3))*$k$)*$k$, 4+2*log($k$)),
    "Reciprocal":  lambda $k$, $\kappa_s$, $\kappa$, $f$, $m$:  
    (24*$k$*$k$+36*$k$, (log($k$)+5)*$k$+5, 
    ((k+8+log($k$))*$k$*3)*$k$, 4+2*log($k$))
    }
\end{lstlisting}

\texttt{Falcon}~\cite{wagh2020falcon}: We configure the communication costs for basic operations of the malicious version of \texttt{Falcon}~\cite{wagh2020falcon} according to Algorithms 5 and 6, Section 3.2, Table 9 and Figure 6 of \cite{wagh2020falcon}. \wqruan{For the communication cost of `share', \texttt{Falcon} uses the same construction as \texttt{ABY3}. For the communication cost of `reveal' according to the description in Section 3.2 of \cite{wagh2020falcon}, each party needs to transfer two messages to the other parties. For the online communication cost of `matmuls', `LTZ', and `Pow2', we directly collect the number of Table 9 of \cite{wagh2020falcon} to configure them. Besides, we configure the communication cost of `muls' operation by setting $p$, $q$, and $r$ of `matmuls' as 1, 1, and 1. For their offline communication costs, `matmuls' does not have the offline phase. Meanwhile, we configure the offline communication cost of `LTZ' according to the description in Figure 6. According to Algorithm 5 of \cite{wagh2020falcon}, `Pow2' is implemented by calling $k$ `LTZ'. Therefore, the offline communication cost of `Pow2' is $k$ times of `LTZ'. For the communication cost of `Reciprocal', according to Algorithm 6 of \cite{wagh2020falcon}, \texttt{Falcon} implements it by calling one `Pow2' and five `muls'. Therefore, we configure the communication cost of `Reciprocal' using a simple calculation. For the communication cost of `TruncPr', \texttt{Falcon} uses the same construction as \texttt{ABY3}. Therefore, we configure its communication cost according to Figure 8 of \cite{mohassel2018aby3}. The communication cost configuration of \texttt{Falcon} is shown in Listing~\ref{list:falcon_cost}.}

\begin{lstlisting}[mathescape,xleftmargin=2em,framexleftmargin=2em, caption = {Communication cost configuration for basic operations of \texttt{Delphi}~\cite{mishra2020delphi}.}, columns=fullflexible, label = {list:delphi_cost}]
class Delphi(Cost):
  cost_func_dict = {
    "share": lambda $k$, $\kappa_s$, $\kappa$, $f$, $m$: (0, 0, 0, 0),
    "reveal": lambda $k$, $\kappa_s$,$\kappa$, $f$, $m$: (2*$k$, 1, 0, 0),
    "muls": lambda $k$, $\kappa_s$,$\kappa$, $f$, $m$, $\mathit{size}$: ($k$*$\mathit{size}$, 1, 
    $\lceil$ $\mathit{size}$/$\mathit{deg}$$\rceil$ *$\mathit{deg}$*sum($\mathit{mod}$)*4, 2),
    "matmuls": lambda $k$, $\kappa_s$, $\kappa$, $f$, $m$, $p$ ,$q$, r, $\mathit{deg}$, $\mathit{mod}$: 
    ($p$*$q$*$k$, 1, ($\lceil$$p$*r/$\mathit{deg}$$\rceil$+$\lceil$$p$*$q$/$\mathit{deg}$$\rceil$) *$\mathit{deg}$*sum($\mathit{mod}$)*2, 2),
    "TruncPr": lambda $k$, $\kappa_s$, $\kappa$, $f$, $m$, $\mathit{knownmsb}$: 
    (0, 0, 0, 0),    
    "LTZ": lambda $k$, $\kappa_s$, $\kappa$, $f$, $m$:  
    (148*$k$, 1, 1470*$k$, 3),
    "conv2d" lambda $k$, $\kappa_s$, $\kappa$, $f$, $m$, $\mathit{batch\_size}$, $\mathit{in\_channel}$, 
    $\mathit{out\_channel}$, $\mathit{inw}$, $\mathit{inh}$, $\mathit{outw}$, $\mathit{outh}$, $\mathit{kw}$, $\mathit{kh}$, $\mathit{deg}$, $\mathit{mod}$: 
    ($\mathit{batch\_size}$*$\mathit{in\_channel}$*$\mathit{inw}$*$\mathit{inh}$*$k$, 1,
    $\mathit{batch\_size}$*$\lceil$$\mathit{in\_channel}$*$\mathit{inw}$*$\mathit{inh}$/$\mathit{deg}$$\rceil$*$\mathit{kw}$*$\mathit{kh}$*$\mathit{deg}$
    *sum($\mathit{mod}$)+$\lceil$$\mathit{batch\_size}$*$\mathit{out\_channel}$*$\mathit{outw}$*$\mathit{outh}$/$\mathit{deg}$$\rceil$
    *$\mathit{deg}$*sum($\mathit{mod}$), 2)
    }
\end{lstlisting}

\wqruan{\texttt{Delphi}~\cite{mishra2020delphi}: We configure the communication cost for basic operations of \texttt{Delphi}~\cite{mishra2020delphi} by analyzing the descriptions in Section 6, Figures 3 and 4 of \cite{mishra2020delphi} and running source codes. Because \texttt{Delphi} has special optimizations for convolution operations, it has a `conv2d'  operation with eight extra parameters: the input channel number $\mathit{in\_channel}$, the output channel number $\mathit{out\_channel}$, the input width $\mathit{inw}$, the input height $\mathit{inh}$, the output width $\mathit{outw}$, the output height $\mathit{outh}$, the kernel width $\mathit{kw}$, the kernel height $\mathit{kh}$. Meanwhile, because \texttt{Delphi} uses additive secret sharing, the communication costs of its `share' and `reveal' operations are consistent with those of \texttt{ABY}.  For the online phase of all linear operations (`muls', `matmuls', `conv2d'), according to Figure 4 of \cite{mishra2020delphi}, their communication costs are the size of input data multiplied by bit length. For the offline phase of linear operations, \texttt{Delphi} directly uses \texttt{GAZELLE}~\cite{juvekar2018gazelle}'s algorithms for linear layers according to the description in Section 6. Therefore, we configure the offline communication costs of \texttt{Delphi} according to the description in Tables II and IV of \texttt{GAZELLE}~\cite{juvekar2018gazelle}. For `LTZ', we configure its online and offline communication cost by testing the source codes of \texttt{Delphi}. The communication cost configuration of \texttt{Delphi} is shown in Listing~\ref{list:delphi_cost}.}

\begin{lstlisting}[mathescape,xleftmargin=2em,framexleftmargin=2em, caption = {Communication cost configuration for basic operations of \texttt{Cheetah}~\cite{Cheetah}. compute\_ct\_num represents the matrix multiplication ciphertext count algorithm defined in Algorithm~\ref{alg:cheetah}}, columns=fullflexible, label = {list:cheetah_cost}]
class Cheetah(Cost):
  cost_func_dict = {
    "share": lambda $k$, $\kappa_s$, $\kappa$, $f$, $m$: (0, 0, 0, 0),
    "reveal": lambda $k$, $\kappa_s$,$\kappa$, $f$, $m$: (2*$k$, 1, 0, 0),
    "muls": lambda $k$, $\kappa_s$,$\kappa$, $f$, $m$, $\mathit{size}$: ($\lceil$$\mathit{size}$/$\mathit{deg}$$\rceil$*
    ($\mathit{deg}$*sum[0:-1]+$\mathit{deg}$*sum[0:-2]), 2, 0, 0),
    "matmuls": lambda $k$, $\kappa_s$, $\kappa$, $f$, $m$, $p$ ,$q$, r, $\mathit{deg}$, $\mathit{mod}$: 
    (2*(comput_ct_num($p$,$q$,r,$\mathit{deg}$)[0]*$\mathit{deg}$*sum($\mathit{mod}$)[0:-1]+
    comput_ct_num($p$,$q$,r,$\mathit{deg}$)[1]*$\mathit{deg}$*sum($\mathit{mod}$)[0:-2]),
    4, 0, 0),
    "TruncPr": lambda $k$, $\kappa_s$, $\kappa$, $f$, $m$, $\mathit{knownmsb}$: 
    ($f$+4, 2, 0, 0),    
    "LTZ": lambda $k$, $\kappa_s$, $\kappa$, $f$, $m$:  
    (13*$k$+1, log($k$), 0, 0)
    }
\end{lstlisting}

\begin{algorithm}[ht]
\small
\caption{Matrix multiplication ciphertext count algorithm of \texttt{Cheetah}~\cite{Cheetah}. }
\label{alg:cheetah}
 \begin{algorithmic}[1]
    \REQUIRE The row number of the first matrix $p$, the column number of the first matrix $q$, the column number of the second matrix r, the degree of HE polynomial $\mathit{deg}$, local computation price $\mathit{lp}$, bandwidth price $\mathit{bp}$
    \ENSURE The number of sending ciphertext $\mathit{s\_ct}$, the number of response ciphertext $\mathit{r\_ct}$
    \STATE \textbf{Initialization}: initialize min\_cost as Integer.MAX, $\mathit{s\_ct}$ = $\mathit{r\_ct}$ =0
    \FOR{$\mathit{d_1}$:$\mathit{min(deg, p+1)}$}
    \STATE $\mathit{block\_num_1}$ = $\lceil $p$/d_1 \rceil$, $\mathit{d_2 = 1}$
    \WHILE{$\mathit{d_2 \leq q}$ and $\mathit{d_1 *d_2} \leq deg$}
    \STATE $\mathit{block\_num_2}$ = $\lceil q/d_2 \rceil$
    \STATE $\mathit{d_3 = min(r, \lceil deg / d_1 / d_2 \rceil)}$
    \STATE $\mathit{block\_num_3}$ = $\lceil r/d_3 \rceil$
    \STATE $\mathit{s\_ct' = min(block\_num_1, block\_num_3)* block\_num_2}$
    \STATE $\mathit{r\_ct' = \lceil block\_num_1*block\_num_3/ d_2 \rceil}$
    \STATE num\_1 = $\mathit{\lceil block\_num_1 * block\_num_3 / deg \rceil*d_2}$
    \STATE num\_2 = $\mathit{block\_num_1}$*$\mathit{block\_num_2}$*$\mathit{block\_num_3}$
    \STATE cost = $(\mathit{s\_ct'}$+$\mathit{r\_ct'}$)*$\mathit{bp}$+ num\_1*$\mathit{lp}$+ num\_2*$\mathit{lp}/10$
    \IF{cost $\leq$ min\_cost}
    \STATE min\_cost = cos
    \STATE $\mathit{s\_ct = s\_ct', r\_ct = r\_ct'}$
    \ENDIF
    \STATE $\mathit{d\_2 = d\_2 *2}$
    \ENDWHILE
    \ENDFOR
    \RETURN $\mathit{s\_ct}$, $\mathit{r\_ct}$
\end{algorithmic}
\end{algorithm}

\wqruan{\texttt{Cheetah}~\cite{Cheetah}:  We configure the communication costs for basic operations of \texttt{Cheetah}~\cite{Cheetah} according to Sections 3.1, 3.2.2, 3.3 and Tables 2, 3, and 4 of \cite{Cheetah}. Besides the above analysis, we further adjust our configuration by the discussion with the developer of the developers of \texttt{SecretFlow}. Because \texttt{Cheetah} uses additive secret sharing,  the communication costs of its `share' and `reveal' operations are consistent with those of \texttt{ABY}. For communication costs of linear operations (`muls', `matmuls'), \texttt{Cheetah} implements them using the CKKS HE mechanism. \texttt{Cheetah} has two optimizations for the usage of the CKKS HE mechanisms. Firstly, the sender who holds the private key of HE can use the symmetric version of CKKS with only half of the ciphertext size. Secondly, according to Section 5.2 of \cite{Cheetah}, the response ciphertext can truncate the low-end parts of the ciphertext. In addition, \texttt{Cheetah} in \texttt{SecretFlow} uses a matrix partition strategy to balance the costs of computation and communication. We show the strategy that is defined in lines 180-223 of `matmat\_prot.cc' of the source codes of \texttt{SecretFlow} in Algorithm~\ref{alg:cheetah}. Combining the above information, we obtain the communication costs of linear operations. For the `LTZ' operation, \texttt{Cheetah} implements it by combining its millionaires' protocol and VOLE-style OT. Therefore, we obtain the communication cost of `LTZ' by collecting the numbers from Tables 2, 3 of \cite{Cheetah}. For the communication cost of `TruncPr', because \texttt{Cheetah} in \texttt{SecretFlow} uses the optimization proposed by Dalskov et al.~\cite{DBLP:journals/popets/Dalskov0K20}, msbs of its truncation input data are always 0. Therefore, we can configure the communication cost of the `TruncPr' according to Table 4 of ~\cite{Cheetah}. The communication cost configuration of \texttt{Cheetah} is shown in Listing~\ref{list:cheetah_cost}.}

\begin{lstlisting}[mathescape,xleftmargin=2em,framexleftmargin=2em, caption = {Communication cost configuration for basic operations of \texttt{Deep-MPC}~\cite{pmlr-v162-keller22a}.}, columns=fullflexible, label = {list:Deep_MPC_cost}]
class Deep-MPC(Cost):
  cost_func_dict = {
    "share": lambda $k$, $\kappa_s$, $\kappa$, $f$, $m$: ($k$, 1, 0, 0),
    "reveal": lambda $k$, $\kappa_s$,$\kappa$, $f$, $m$: (3*$k$, 1, 0, 0),
    "muls": lambda $k$, $\kappa_s$,$\kappa$, $f$, $m$, $\mathit{size}$: (3*$k$, 1, 0, 0),
    "matmuls": lambda $k$, $\kappa_s$, $\kappa$, $f$, $m$, $p$ ,$q$, r, $\mathit{deg}$, $\mathit{mod}$: 
    (3*$p$*r*$k$, 1, 0, 0),
    "TruncPr": lambda $k$, $\kappa_s$, $\kappa$, $f$, $m$, $\mathit{knownmsb}$: 
    (8*$k$, 3, 0, 0),    
    "LTZ": lambda $k$, $\kappa_s$, $\kappa$, $f$, $m$:  
    (7.425*$k$, log($k$)+2, 3*$k$, 2)
    }
\end{lstlisting}

\wqruan{\texttt{Deep-MPC}~\cite{pmlr-v162-keller22a}: We configure the communication costs for basic operations of the three-party protocol of \texttt{Deep-MPC}~\cite{pmlr-v162-keller22a} by analyzing Appendix A, Table 8 of \cite{pmlr-v162-keller22a} and running its source codes in \texttt{MP-SPDZ}. Keller and Sun listed communication costs of `reveal', `muls', `TruncPr', `LTZ' in Table 8 of \cite{pmlr-v162-keller22a} of \texttt{Deep-MPC}. Therefore, we directly collect numbers listed in Table 8 of \cite{pmlr-v162-keller22a} to configure the communication costs of the three basic operations. Meanwhile, we run the source codes of \texttt{MP-SPDZ} and align the communication cost of `TruncPr' to the newest implementation of \texttt{Deep-MPC}. Besides, as stated in Appendix A of \cite{pmlr-v162-keller22a}, \texttt{Deep-MPC} uses dabits~\cite{10.1007/978-3-030-35423-7_12} to convert boolean sharing to arithmetic sharing. Therefore, the cost for `LTZ' in Table 8 of \cite{pmlr-v162-keller22a} is split into online and offline communication costs (7.425k bits for online and 3k bits for offline). For the `matmuls' operation, as stated in Appendix A of \cite{pmlr-v162-keller22a}, \texttt{Deep-MPC} uses the same construction as \texttt{ABY3}, the communication cost of `matmuls' is the same as that of \texttt{ABY3}. For the `reveal' operation, as stated in Appendix A of \cite{pmlr-v162-keller22a}, \texttt{Deep-MPC} employs the optimization proposed by Eerikson et al.~\cite{DBLP:conf/icits/EeriksonKOPP020}. Therefore, the `reveal' operation of \texttt{Deep-MPC} only requires one message transfer. The communication cost configuration of \texttt{Deep-MPC} is shown in Listing~\ref{list:Deep_MPC_cost}.}

\begin{lstlisting}[mathescape,xleftmargin=2em,framexleftmargin=2em, caption = {Communication cost configuration for basic operations of \texttt{SecretFlow-SEMI2K}~\cite{secretflow}. compute\_mmul\_size represents the matrix multiplication message count algorithm defined in Algorithm~\ref{alg:secretFlow-SEMI2K}}, columns=fullflexible, label = {list:secretflow_semi2k_cost}]
class SecretFlow-SEMI2K(Cost):
  cost_func_dict = {
    "share": lambda $k$, $\kappa_s$, $\kappa$, $f$, $m$: (0, 0, 0, 0),
    "reveal": lambda $k$, $\kappa_s$,$\kappa$, $f$, $m$: ($m$*($m$-1)*$k$, 1, 0, 0),
    "muls": lambda $k$, $\kappa_s$,$\kappa$, $f$, $m$, $\mathit{size}$: (2*$m$*($m$-1)*$k$*$\mathit{size}$, 1, 0, 0),
    "matmuls": lambda $k$, $\kappa_s$, $\kappa$, $f$, $m$, $\mathit{mod}$ ,$q$, r, $\mathit{deg}$, $\mathit{mod}$: 
    ($m$*$k$*compute_mmul_size($\mathit{mod}$, $q$, r, $k$), 1,0, 0),
    "TruncPr": lambda $k$, $\kappa_s$, $\kappa$, $f$, $m$, $\mathit{knownmsb}$: 
    ($m$*($m$-1)*$k$, 1, 0, 0),    
    "LTZ": lambda $k$, $\kappa_s$, $\kappa$, $f$, $m$:  
    ($m$*(2*$k$ + 2*($m$-1)*(2*$k$+32)), math.log2($k$)+1
    , 0, 0)
    }
\end{lstlisting}

\begin{algorithm}[ht]
\small
\caption{Matrix multiplication message count algorithm of \texttt{SecretFlow-SEMI2K}~\cite{secretflow}. }
\label{alg:secretFlow-SEMI2K}
 \begin{algorithmic}[1]
    \REQUIRE The row number of the first matrix $p$, the column number of the first matrix $q$, the column number of the second matrix r, bit length $k$
    \ENSURE The number of ring elements required for the input matrices multiplication $num\_e$
    \STATE \textbf{Initialization}: initialize $\mathit{mem\_limit}$ as $2^{31}$, $\mathit{res} =0$
    \IF{ $p=0 \; or \; q=0 \; or\; r=0 \; or$ $\mathit{(p*q+q*r)*k <}$ $\mathit{mem\_limit}$}
    \RETURN $\mathit{p*q+q*r}$
    \ENDIF
    \STATE  q\_step = expected\_pr\_step =0
    \IF{$\mathit{q > (p+r)*8}$}
    \STATE expected\_pr\_step = $\mathit{p+r}$
    \STATE q\_step = max(1, $\lceil$ $\mathit{mem\_limit}$/$k$/ expected\_pr\_step$\rceil$)
    \ELSIF{$\mathit{(p+r) > q*8}$}
    \STATE  q\_step = $q$
    \STATE expected\_pr\_step = max(1, $\lceil$ $\mathit{mem\_limit}/k/q\_step \rceil$)
    \ELSE
    \STATE pr\_step = $\sqrt{\mathit{(p+r)*mem\_limit}/(q*k)}$
    \STATE q\_step = max(1, $\lceil \mathit{mem\_limit/(k)}$/pr\_step $\rceil$)
    \STATE expected\_pr\_step = max(1, $\lceil$ pr\_step $\rceil$)
    \ENDIF
    \STATE p\_step = max(1, $\lceil$ expected\_pr\_step*$\mathit{mod}$/($\mathit{mod}$+r) $\rceil$)
    \STATE r\_step = max(1, $\lceil$ expected\_pr\_step*r/($\mathit{mod}$+r) $\rceil$)
    \FOR{$\mathit{i}$:$\mathit{\lceil p/}$ p\_step $\rceil$}
    \FOR{$\mathit{j}$:$\mathit{\lceil q/}$ q\_step $\rceil$}
    \FOR{$\mathit{k}$:$\mathit{\lceil r/}$ r\_step $\rceil$}
    \STATE p\_sub = min($\mathit{mod}$ - p\_step*$i$, p\_step)
    \STATE q\_sub = min(q - q\_step*$i$, q\_step)
    \STATE r\_sub = min(r - r\_step*$i$, r\_step)
    \STATE $\mathit{res}$ += p\_sub*q\_sub + q\_sub*r\_sub
    \ENDFOR
    \ENDFOR
    \ENDFOR
    \RETURN $\mathit{res}$
\end{algorithmic}
\end{algorithm}

\wqruan{\texttt{SecretFlow-SEMI2K}~\cite{secretflow}: We configure the communication costs for basic operations of \texttt{SecretFlow-SEMI2K}~\cite{secretflow} according to communication costs of linear operations (`share', `reveal', `muls', `matmuls', `TruncPr') and the comparison operation (`LTZ') listed in `arithmetic.h' and `conversion.h' of `libspu/mpc/semi2k/'. Besides, \texttt{SecretFlow-SEMI2K} partitions the input matrices of matrix multiplication to avoid memory overflow. We show the message count algorithm for matrix multiplication in Algorithm~\ref{alg:secretFlow-SEMI2K} according to the matrix partition strategy defined in lines 249-289 of `ring.cc' of `libs/kernel/hal/'. The communication cost configuration of \texttt{SecretFlow-SEMI2K} is shown in Listing~\ref{list:secretflow_semi2k_cost}. Note that we configure the communication costs of \texttt{SecretFlow-SEMI2K} for its trusted first-party (TFP) setting. Therefore, the offline communication costs of basic operations listed in Listing~\ref{list:secretflow_semi2k_cost} are all zero.}

\begin{table*}[ht]
\caption{\wqruan{The online communication size profiling results of two secure CNN model inference processes outputted by \texttt{Delphi}~\cite{mishra2020delphi} and \hawkeye.} }
\centering
\scalebox{0.7}{
\begin{tabular}{c|c|ccc|c}
\hline
Model                      & Framework                        & \multicolumn{3}{c|}{\% (MB) of linear operators}                                                                                                                                                                                                                                  & \% (MB) of non-linear operators                                                \\ \hline
\multicolumn{1}{l|}{}      &                                  & \multicolumn{1}{c|}{\% (MB) of Conv2d}                                                           & \multicolumn{1}{c|}{\% (MB) of other linear operators}                                           & \% (MB) of all                                                              & \multicolumn{1}{l}{}                                                           \\ \hline
\multirow{3}{*}{MiniONN}   & \texttt{Delphi}~\cite{mishra2020delphi} & \multicolumn{1}{c|}{\wqruan{0.50\% (0.98MB)}}                                                             & \multicolumn{1}{c|}{\wqruan{0.01\% (0.01MB)}}                                                             & \wqruan{0.51\% (0.99MB)}                                                             & \wqruan{99.49\% (195.41MB)}                                                             \\ \cline{2-6} 
                           & \hawkeye          & \multicolumn{1}{c|}{\begin{tabular}[c]{@{}c@{}}\wqruan{0.44\% (0.87MB)}\\ \wqruan{-0.06\% (-0.11MB)}\end{tabular}} & \multicolumn{1}{c|}{\begin{tabular}[c]{@{}c@{}}\wqruan{0.01\% (0.01MB)}\\ \wqruan{-0.00\% (-0.00MB)}\end{tabular}} & \begin{tabular}[c]{@{}c@{}}\wqruan{0.45\% (0.88MB)}\\ \wqruan{-0.06\% (-0.11MB)}\end{tabular} & \begin{tabular}[c]{@{}c@{}}\wqruan{99.55\% (195.41MB)}\\ \wqruan{+0.06\% (+0.00MB)}\end{tabular}      \\ \hline
\multirow{3}{*}{ResNet-32} & \texttt{Delphi}~\cite{mishra2020delphi} & \multicolumn{1}{c|}{\wqruan{0.75\% (2.59MB)}}                                                             & \multicolumn{1}{c|}{\wqruan{0.00\% (0.01MB)}}                                                             & \wqruan{0.75\% (2.60MB)}                                                             & \wqruan{99.25\% (342.25MB)}                                                             \\ \cline{2-6} 
                           & \hawkeye          & \multicolumn{1}{c|}{\begin{tabular}[c]{@{}c@{}}\wqruan{0.73\% (2.52MB)}\\ \wqruan{-0.02\% (-0.07MB)}\end{tabular}} & \multicolumn{1}{c|}{\begin{tabular}[c]{@{}c@{}}\wqruan{0.00\% (0.01MB)}\\ \wqruan{+0.00\% (-0.00MB)}\end{tabular}} & \begin{tabular}[c]{@{}c@{}}\wqruan{0.73\% (2.53MB)}\\ \wqruan{-0.02\% (-0.07MB)}\end{tabular} & \begin{tabular}[c]{@{}c@{}}\wqruan{99.27\% (342.25MB)}\\ \wqruan{+0.02\% (+0.00MB)}\end{tabular} \\ \hline
\end{tabular}}
\label{tab:delphi_online}

\end{table*}

\begin{table*}[ht]
\caption{\wqruan{The offline communication size profiling results of two secure CNN model inference processes outputted by \texttt{Delphi}~\cite{mishra2020delphi} and \hawkeye.}}
\centering
\scalebox{0.7}{
\begin{tabular}{c|c|ccc|c}
\hline
Model                      & Framework                        & \multicolumn{3}{c|}{\% (MB) of linear operators}                                                                                                                                                                                                                                        & \% (MB) of non-linear operators                                                 \\ \hline
\multicolumn{1}{l|}{}      &                                  & \multicolumn{1}{c|}{\% (MB) of Conv2d}                                                              & \multicolumn{1}{c|}{\% (MB) of other linear operators}                                           & \% (MB) of all                                                                 & \multicolumn{1}{l}{}                                                            \\ \hline
\multirow{3}{*}{MiniONN}   & \texttt{Delphi}~\cite{mishra2020delphi} & \multicolumn{1}{c|}{\wqruan{2.34\% (75.52MB)}}                                                               & \multicolumn{1}{c|}{\wqruan{0.03\% (1.00MB)}}                                                             & \wqruan{2.37\% (76.52MB)}                                                               & \wqruan{97.63\% (3150.27MB)}                                                             \\ \cline{2-6} 
                           & \hawkeye          & \multicolumn{1}{c|}{\begin{tabular}[c]{@{}c@{}}\wqruan{2.00\% (64.29MB)}\\ \wqruan{-0.34\% (-11.23MB)}\end{tabular}}  & \multicolumn{1}{c|}{\begin{tabular}[c]{@{}c@{}}\wqruan{0.03\% (0.85MB)}\\ \wqruan{+0.00\% (-0.15MB)}\end{tabular}} & \begin{tabular}[c]{@{}c@{}}\wqruan{2.03\% (65.14MB)}\\ \wqruan{-0.34\% (-11.38MB)}\end{tabular}  & \begin{tabular}[c]{@{}c@{}}\wqruan{97.97\% (3148.73MB)}\\ \wqruan{+0.34\% (-1.54MB)}\end{tabular} \\ \hline
\multirow{3}{*}{ResNet-32} & \texttt{Delphi}~\cite{mishra2020delphi} & \multicolumn{1}{c|}{\wqruan{3.70\% (212.04MB)}}                                                              & \multicolumn{1}{c|}{\wqruan{0.02\% (1.00MB)}}                                                             & \wqruan{3.72\% (213.04MB)}                                                              & \wqruan{96.28\% (5517.63MB)}                                                             \\ \cline{2-6} 
                           & \hawkeye          & \multicolumn{1}{c|}{\begin{tabular}[c]{@{}c@{}}\wqruan{3.43\% (195.86MB)}\\ \wqruan{-0.27\% (-16.18MB)}\end{tabular}} & \multicolumn{1}{c|}{\begin{tabular}[c]{@{}c@{}}\wqruan{0.01\% (0.85MB)}\\ \wqruan{-0.01\% (-0.15MB)}\end{tabular}} & \begin{tabular}[c]{@{}c@{}}\wqruan{3.44\% (196.71MB)}\\ \wqruan{-0.28\% (-16.33MB)}\end{tabular} & \begin{tabular}[c]{@{}c@{}}\wqruan{96.56\% (5516.39MB)}\\ \wqruan{+0.28\% (-1.24MB)}\end{tabular} \\ \hline
\end{tabular}}
\label{tab:delphi_offline}
\end{table*}

\section{Accuracy of \hawkeye on Mixed-protocol Frameworks}~\label{appendix:mixed-protocol}

\wqruan{\noindent \textbf{Experiment Setup.} In order to further show that \hawkeye works well with mixed-protocol frameworks that rely on HE or GC, we compare the communication cost profiling results outputted by \hawkeye and two MPL frameworks (i.e. \texttt{Delphi}~\cite{mishra2020delphi} and \texttt{Cheetah}~\cite{Cheetah}) that rely on HE and GC respectively. We run two secure CNN models (MiniONN~\cite{minionn} and ResNet-32~\cite{7780459}) inference processes on the open-source codes~\footnote{The code repository address of \texttt{Delphi} is \url{https://github.com/mc2-project/delphi}. We use codes at commit \textit{92bc007}}~\footnote{The code repository address of \texttt{Cheetah} is \url{https://github.com/secretflow/spu}. We use \texttt{SecretFlow-SPU0.9.3b0}} of these two MPL frameworks to obtain the dynamic profiling results. We choose MiniONN~\cite{minionn} and ResNet-32~\cite{7780459} because these two models are tested in the paper of \texttt{Delphi} and \texttt{Cheetah} simultaneously. Because \texttt{Delphi} merges the batch normalization layer and the convolution layer into a single convolution operation, we omit the batch normalization layer of these two models.  For parameter setting, we set the bit length as 64 and the bit length of fixed numbers' fractional part as 18. We set the statistical security parameter and computation security parameter of these two MPL frameworks as $40$ and $128$, respectively. Besides, we set the polynomial degree and modulus coefficients of BFV HE used by \texttt{Delphi} as 8192 and \{43, 43, 44, 44, 44\} respectively. For the CKKS HE used by \texttt{Cheetah}, we set the polynomial degree and modulus coefficients as 8192 and \{59, 55, 49, 49\} respectively. For the input data, following the setting of \texttt{Delphi}, we set the input data size as $1 \times 3 \times 32 \times 32$.}

\wqruan{To obtain the static model communication cost profiling results from \hawkeye, we configure the model communication cost of \texttt{Delphi} and \texttt{Cheetah} in \hawkeye and implement the above models. Finally, we run \hawkeye under the same parameter setting with \texttt{Delphi} and \texttt{Cheetah} to obtain the static profiling results. We also show that communication cost configuration of \texttt{Delphi} and \texttt{Cheetah} in Appendix~\ref{appendix:protocol_config}.}

\wqruan{\textbf{Experimental results.} As is shown in Table~\ref{tab:delphi_online}, Table~\ref{tab:delphi_offline}, and Table~\ref{tab:cheetah}, \hawkeye can accurately profile the model communication cost on \texttt{Delphi} and \texttt{Cheetah}. Concretely, the proportions of operators' online/offline communication size to the total online/offline communication size outputted by \hawkeye and baselines are very close. The differences between the communication size profiling results outputted by \hawkeye and baselines could be caused by the following two reasons: (1) The HE packing is usually not compact in the implementation. \texttt{Delphi} and \texttt{Cheetah} both use HE to compute linear operators. HE packs multiple plaintext data into one ciphertext to reduce the communication overhead. However, due to implementation issues, the HE packing usually cannot be as compact as the theoretical. Therefore, the communication sizes of linear operators outputted by \texttt{Delphi} and \texttt{Cheetah} are slightly larger than those outputted by \hawkeye (2) The complexity of the comparison operations is asymptotic rather than actual. Similar to \texttt{CrypTFlow2}, \texttt{Cheetah} only provides an upper bound of the communication complexity for its comparison operators. Therefore, the communication sizes of non-linear operators outputted by \hawkeye are slightly larger than those outputted by \texttt{Cheetah}. }

\begin{table*}[ht]
\caption{\wqruan{The online communication size profiling results of outputted by \texttt{Cheetah}~\cite{Cheetah} and \hawkeye for two secure CNN model inference processes.}}
\centering
\scalebox{0.7}{
\begin{tabular}{c|c|ccc|c}
\hline
Model                      & Framework                         & \multicolumn{3}{c|}{\% (MB) of linear operators}                                                                                                                                                                                                                                      & \% (MB) of non-linear operators                                               \\ \hline
\multicolumn{1}{l|}{}      &                                   & \multicolumn{1}{c|}{\% (MB) of Conv2d}                                                             & \multicolumn{1}{c|}{\% (MB) of other linear operators}                                           & \% (MB) of all                                                                & \multicolumn{1}{l}{}                                                          \\ \hline
\multirow{3}{*}{MiniONN}   & \texttt{Cheetah}~\cite{Cheetah} & \multicolumn{1}{c|}{\wqruan{53.99\% (20.14MB)}}                                                             & \multicolumn{1}{c|}{\wqruan{2.41\% (0.90MB)}}                                                             & \wqruan{56.40\% (21.04MB)}                                                             & \wqruan{43.60\% (16.26MB)}                                                             \\ \cline{2-6} 
                           & \hawkeye           & \multicolumn{1}{c|}{\begin{tabular}[c]{@{}c@{}}\wqruan{50.47\% (18.34MB)}\\ \wqruan{-3.52\% (-1.80MB)}\end{tabular}}  & \multicolumn{1}{c|}{\begin{tabular}[c]{@{}c@{}}\wqruan{2.26\% (0.82MB)}\\ \wqruan{-0.15\% (-0.08MB)}\end{tabular}} & \begin{tabular}[c]{@{}c@{}}\wqruan{52.73\% (19.16MB)}\\ \wqruan{-3.67\% (-1.88MB)}\end{tabular} & \begin{tabular}[c]{@{}c@{}}\wqruan{47.27\% (17.18MB)}\\ \wqruan{+3.67\% (+0.92MB)}\end{tabular} \\ \hline
\multirow{3}{*}{ResNet-32} & \texttt{Cheetah}~\cite{Cheetah} & \multicolumn{1}{c|}{\wqruan{64.40\% (53.03MB)}}                                                             & \multicolumn{1}{c|}{\wqruan{1.02\% (0.84MB)}}                                                             & \wqruan{65.42\% (53.87MB)}                                                             & \wqruan{34.58\% (28.47MB)}                                                             \\ \cline{2-6} 
                           & \hawkeye           & \multicolumn{1}{c|}{\begin{tabular}[c]{@{}c@{}}\wqruan{63.14\% (52.87MB)}\\ \wqruan{-1.26\% (-0.16MB)}\end{tabular}} & \multicolumn{1}{c|}{\begin{tabular}[c]{@{}c@{}}\wqruan{0.92\% (0.77MB)}\\ \wqruan{-0.10\% (-0.07MB)}\end{tabular}} & \begin{tabular}[c]{@{}c@{}}\wqruan{64.06\% (53.64MB)}\\ \wqruan{-1.36\% (-0.23MB)}\end{tabular} & \begin{tabular}[c]{@{}c@{}}\wqruan{35.94\% (30.10MB)}\\ \wqruan{+1.36\% (+1.63MB)}\end{tabular} \\ \hline
\end{tabular}}
\label{tab:cheetah}
\end{table*}
\begin{table*}[ht]
\caption{\wqruan{The online communication size profiling results of two secure CNN model training processes outputted by \texttt{SecreFLow-SEMI2K}~\cite{secretflow} and \hawkeye.}}
\centering
\scalebox{0.7}{
\begin{tabular}{c|c|c|c|c}
\hline
Model                      & Framework                                                                           & \% (GB) of Model Forward                                                             & \% (GB) of Model Backward                                                            & \% (GB) of Model Update                                                           \\ \hline
\multirow{3}{*}{ResNet-50}    & \texttt{SecretFlow-SEMI2K}~\cite{secretflow} & \wqruan{45.59\% (120.75GB)}                                                             & \wqruan{54.28\% (143.78GB)}                                                             & \wqruan{0.13\% (0.35GB)}                                                             \\ \cline{2-5} 
                           & \hawkeye                                                             & \begin{tabular}[c]{@{}c@{}}\wqruan{44.02\% (120.04GB)}\\ \wqruan{-1.57\% (-0.71GB)}\end{tabular} & \begin{tabular}[c]{@{}c@{}}\wqruan{55.85\% (152.29GB)}\\ \wqruan{+1.57\% (+8.51GB)}\end{tabular} & \begin{tabular}[c]{@{}c@{}}\wqruan{0.13\% (0.35GB)}\\ \wqruan{+0.00\% (+0.00GB)}\end{tabular} \\ \hline
\multirow{3}{*}{VGG-16} & \texttt{SecretFlow-SEMI2K}~\cite{secretflow} & \wqruan{47.49\% (16.92GB)}                                                              & \wqruan{51.89\% (18.49GB)}                                                              & \wqruan{0.62\% (0.22GB)}                                                             \\ \cline{2-5} 
                           & \hawkeye                                                             & \begin{tabular}[c]{@{}c@{}}\wqruan{45.69\% (16.77GB)}\\ \wqruan{-1.80\% (-0.15GB)}\end{tabular}  & \begin{tabular}[c]{@{}c@{}}\wqruan{53.71\% (19.71GB)}\\ \wqruan{+1.82\% (+1.22GB)}\end{tabular}  & \begin{tabular}[c]{@{}c@{}}\wqruan{0.60\% (0.22GB)}\\ \wqruan{-0.02\% (+0.00GB)}\end{tabular} \\ \hline
\end{tabular}}
\label{tab:semi2k_training}
\end{table*}

\section{Accuracy of \hawkeye on Secure Model Training Scenarios}~\label{appdendix:secure_training}

\wqruan{\textbf{Experiment Setup.} To show that \hawkeye works well in secure model training scenarios, we compare the communication cost profiling results outputted by \hawkeye and \texttt{SecretFLow-SEMI2K}, the two-party backend of \texttt{SecretFlow-SPU}~\cite{secretflow}. We use classical VGG-16 and ResNet-50 models as our target models. To obtain the dynamic communication cost profiling results, we run secure VGG-16 and ResNet-50 model training processes on \texttt{SecretFLow-SEMI2K}~\footnote{The code repository address of \texttt{SecretFLow-SEMI2K} is \url{https://github.com/secretflow/spu}. We use \texttt{SecretFlow-SPU0.9.3b0}}.  For parameter setting, we set the bit length as 64 and the bit length of fixed numbers' fractional part as 18. We set the statistical security parameter and computation security parameter of these two MPL frameworks as $40$ and $128$, respectively. Meanwhile, We use SGD as the optimizer. The input data size is $128\times3\times32\times32$, where 128 is the batch size.}

\wqruan{To obtain the static model communication cost profiling results from \hawkeye, we configure the model communication cost of \texttt{SecretFLow-SEMI2K} in \hawkeye and implement the above models. Finally, we run \hawkeye under the same parameter setting with \texttt{SecretFLow-SEMI2K} to obtain the static profiling results. We show the communication cost configuration of \texttt{SecretFLow-SEMI2K} in Appendix~\ref{appendix:protocol_config}.}

\wqruan{\textbf{Experimental results.} As is shown in Table~\ref{tab:semi2k_training}, \hawkeye can accurately profile the communication cost of the secure model training process. The proportions of model training phases' online communication size to the total online communication size outputted by \texttt{SecretFlow-SEMI2K} and \hawkeye are very close. Concretely, the proportion differences between baselines and \hawkeye are all smaller than 1.82\%. The main difference lies in the model backward process. The potential reason behind it is that \texttt{SecretFlow-SEMI2K} has optimizations for a few special circuit structures, such as consecutive multiplications used in the model backward process. Therefore, the online communication sizes of the model backward process outputted by \texttt{SecretFlow-SEMI2K} are slightly smaller than those outputted by \hawkeye.}